\pdfoutput=1
\documentclass[12pt]{article}
\usepackage{graphicx,amssymb,amsmath}
\usepackage{epsfig}
\usepackage{color,graphicx}
\usepackage{cite}
\usepackage{amssymb,amsmath}
\usepackage{multirow}
\usepackage{hyperref}
\newcommand\fverb{\setbox\pippobox=\hbox\bgroup\verb}
\newcommand\fverbdo{\egroup\medskip\noindent%
                              \fbox{\unhbox\pippobox}\ }
\newcommand\fverbit{\egroup\item[\fbox{\unhbox\pippobox}]}
\newbox\pippobox

\newcommand{\beq} {\begin{equation}}
\newcommand{\eeq} {\end{equation}}
\newcommand{\beqa} {\begin{eqnarray}}
\newcommand{\eeqa} {\end{eqnarray}}

\newcommand{\eps}{\epsilon}

\newcommand{\la}{\lambda}

\newcommand{\be}{\begin{equation}}
\newcommand{\ee}{\end{equation}}
\newcommand{\bea}{\begin{eqnarray}}
\newcommand{\eea}{\end{eqnarray}}

\def\la{\lambda}
\def\l{\lambda}
\def\t{\tau}

\begin{document}
 
\begin{flushright}
HIP-2018-18/TH
\end{flushright}

\begin{center}

\centering{\Large {\bf Holographic QCD in the Veneziano limit\\[3mm] and neutron stars}}

\vspace{8mm}

\renewcommand\thefootnote{\mbox{$\fnsymbol{footnote}$}}
Niko Jokela,${}^{1,2}$\footnote{niko.jokela@helsinki.fi}
Matti J\"arvinen,${}^3$\footnote{m.o.jarvinen@uu.nl} and
Jere Remes${}^{1,2}$\footnote{jere.remes@helsinki.fi}

\vspace{4mm}
${}^1${\small \sl Department of Physics} and ${}^2${\small \sl Helsinki Institute of Physics} \\
{\small \sl P.O.Box 64} \\
{\small \sl FIN-00014 University of Helsinki, Finland}

 \vspace{2mm}
 \vskip 0.2cm
 ${}^3${\small \sl Institute for Theoretical Physics} \\
 {\small \sl Utrecht University} \\
 {\small \sl Princetonplein 5, 3584 CC Utrecht, The Netherlands} 

\end{center}

\vspace{8mm}

\setcounter{footnote}{0}
\renewcommand\thefootnote{\mbox{\arabic{footnote}}}

\begin{abstract}
\noindent 
We use the holographic V-QCD models to analyse the physics of dense QCD and neutron stars. Accommodating lattice results for thermodynamics of QCD enables us to make generic predictions for the Equation of State (EoS) of the quark matter phase in the cold and dense regime. We demonstrate that the resulting pressure in V-QCD matches well with a family of neutron-star-matter EoSs that interpolate between state-of-the-art theoretical results for low and high density QCD. After implementing the astrophysical constraints, \emph{i.e.}, the largest known neutron star mass and the recent LIGO/Virgo results for the tidal deformability, we analyse the phase transition between the baryonic and quark matter phases.  We find that the baryon density $n_B$ at the transition is at least 2.9 times the nuclear saturation density $n_s$. The transition is of strongly first order at low and intermediate densities, \emph{i.e.}, for $n_B/n_s \lesssim 7.5$. 
\end{abstract}

\newpage
\tableofcontents


\newpage


\section{Introduction}
\label{sec:introduction}

Last year witnessed a remarkable achievement that marked the beginning era of multimessenger astrophysics: the observation of a neutron star collision by the LIGO and Virgo collaborations.  The aftermath of analysing the gravitational wave data and the associated electromagnetic kilonova observations still continues. A particularly thrilling fact is that already this first observation of a neutron star merger sets bounds on the properties of the underlying theory, i.e., quantum chromodynamics (QCD) at high densities and low temperatures. In the future more neutron star mergers are to be observed, which will then dictate how matter will behave under extreme conditions as quantified in the Equation of State (EoS). 

Connecting these observations with the theory is, however, hard because QCD cannot be solved in the relevant regime.
The cores of cold neutron stars are extremely dense so that they cannot be described in terms of weakly interacting baryons, but not dense enough for perturbative QCD (pQCD) to be applicable.
 Lattice calculations are not available in this regime either, and consequently it is not even known what the relevant  microscopic degrees of freedom are at the center of a typical neutron star. 

The gauge/gravity duality may provide us with a way forward. The applications of the gauge/gravity duality to quark-gluon plasma has been a very active field \cite{CasalderreySolana:2011us,Brambilla:2014jmp}. Very recently these techniques have also been applied in the context of neutron stars \cite{Hoyos:2016zke,Annala:2017tqz} and stiff phases of matter \cite{Hoyos:2016cob,Ecker:2017fyh} which are likely to be needed to achieve large enough stars \cite{Bedaque:2014sqa}. In \cite{Hoyos:2016zke}, a holographic model dual to $\mathcal{N}=4$ super-Yang-Mills and quenched supersymmetric matter was studied, and it was observed that neutron stars can only accommodate very small amounts of quark matter.

In this article, we aim at progress in predicting the EoS of dense QCD matter by using more realistic holographic models, which break conformal symmetry, include confinement, and are able to model the dynamics of real QCD with good precision. The class of models that we will adopt is called V-QCD, which are effective, bottom-up holographic models for QCD~\cite{jk}. The letter ``V'' in the acronym stands for the Veneziano limit: we take $N_f,N_c\to \infty$ while keeping $x_f=N_f/N_c$ fixed. This means that unlike most of the holographic models for QCD considered in the literature, it contains a fully dynamical quark sector on top of the gluons. The gluon sector is given by improved holographic QCD~\cite{ihqcd1,ihqcd2,ihqcd3}, {\emph{i.e.}}, five dimensional Einstein-dilaton gravity with a properly chosen dilaton potential, whereas the quark sector is given by tachyonic DBI (Dirac-Born-Infeld) actions~\cite{Bigazzi:2005md,ckp,ikp1,ikp2}. Naturally, taking the backreaction of the quarks into account is particularly important at high densities. 

The description of dynamical quarks comes with a price to pay: these models are also clearly more complicated than typical holographic models of QCD. Being effective models, the actions contain several \emph{a priori} unknown potentials which need to be chosen by comparing the model predictions to QCD physics. With a correct choice of action the models have known qualitative features of QCD such as asymptotic freedom, linear confinement, and discrete spectrum of hadrons and glueballs~\cite{ihqcd1,ihqcd2,jk,aijk1,aijk2}. The gluon sector has also been shown to agree well with lattice thermodynamics for pure Yang-Mills theory and with lattice results for the glueball masses~\cite{ihqcd6}.

In this article, we take the first steps of comparing the full, backreacted V-QCD models directly to lattice QCD data. Specifically we choose a set of models which fit well the three-flavor lattice EoS at $\mu=0$ and its first nontrivial cumulant around $\mu=0$. Remarkably, we show that a very good fit of the lattice data is possible by using relatively simple Ans\"atze for the potentials. We demonstrate that already after this fit, the EoS at low temperatures and high densities is realistic, and constrained enough to make nontrivial predictions.

We then use the holographic method to model the EoSs in the deconfined quark matter phase, and a set of interpolated EoSs (see, {\emph{e.g.}},~\cite{Annala:2017llu}), which are constrained by matching with effective theories for nuclear matter at low densities and with pQCD at high densities, to model the equation state in the baryonic phase. This construction results in a family of all possible EoSs which are consistent with the holographic model and our knowledge of QCD.  Further demanding that neutron stars (obtained by solving the TOV equations) satisfy the known tidal deformability constraints and are able to support two-solar-mass stars enables us to make generic predictions.  In particular, we consider the possibility of finding
quark matter  deep in the cores of neutron stars. We find the answer to be negative. Even though this means that the state of the neutron star is determined by the interpolated EoSs rather than holography, we can still make nontrivial predictions on the location and nature of the phase transition between the hadronic and quark matter phases. Most importantly, we can set relatively tight constraints on the latent heat at the transition, which is an important parameter in the neutron star mergers \cite{CheslerJokelaLoebVuorinen}. The basic picture which arises is the following: the nuclear (holographic) matter has a stiff (soft) equation of state, and the latent heat at the transition between the phases is sizeable.

The article is organized as follows. In the following section we will review the holographic model. In Section \ref{sec:fit} we will determine the forms of the potentials in the V-QCD action by fitting them to available lattice QCD data. Section \ref{sec:thermo} discusses the thermodynamics of the holographic model at finite density and vanishing temperature. We also explain how we determine that the resulting pressures are physically viable by comparing them to a family of polytropic equations of state that are thermodynamically consistent and have the correct asymptotic behavior both at high and low density. In Section \ref{sec:hybrids} we construct hybrid equations of state where the baryonic EoS at low density, given in terms of the polytropes, has a single transition to the quark matter phase, which is modeled by our holographic framework. The EoSs are then confronted with the astrophysical constraints, and predictions for the phase transition and neutron star masses are derived. Finally, Section \ref{sec:discussion} both contains  the discussion and outlines some thoughts about future outgrowths of our work. Details of the fitting of the potentials are relegated in Appendix \ref{app:fit}.

\section{Holographic model}

In this work we will be using a class of holographic models for QCD (V-QCD)~\cite{jk}. These models are effective bottom-up models containing both gluons and dynamical quarks which are fully backreacted to the glue in the Veneziano limit: 
\be
 N_c \to \infty\, \quad\mathrm{and}\quad \ N_f \to \infty  \quad \mathrm{with} \quad x_f=\frac{N_f}{N_c} \, \quad\mathrm{and}\quad g^2 N_c \quad \mathrm{fixed} \,.
\ee
The glue sector is described in terms of improved holographic QCD (IHQCD), \emph{i.e.} five dimensional Einstein-dilaton gravity with properly tuned dilaton potential~\cite{ihqcd1,ihqcd2}. The quark sector is described by a tachyonic DBI + CS (Chern-Simons) action which is associated to a pair of space filling $D4$--$\overline{D4}$ branes~\cite{Bigazzi:2005md,ckp}. 

The dictionary is as follows:
\begin{itemize}
 \item The dilaton $\l = e^\phi$ is dual to the $\mathrm{Tr} F^2$ operator and therefore sources the 't Hooft coupling $g^2 N_c$ of the Yang-Mills theory.
 \item The tachyon $\t$ is dual to the $\bar qq $ operator and sources the quark mass. In this article, we will be considering chirally symmetric backgrounds so that the tachyon vanishes.
 \item The (vectorial) gauge field $A_\mu$ is dual to the current $J_\mu = \bar q \gamma_\mu q$. The temporal component therefore sources the quark chemical potential.
\end{itemize}
Here we already suppressed the flavor structure -- our solutions will be flavor independent since all quark masses vanish.

The terms in the action of the model which are relevant for the current study are
\be
 S_\mathrm{V-QCD} = S_g + S_f
\ee
where
\begin{align}\label{Sg}
 S_g &=  M^3 N_c^2 \int d^5x \ \sqrt{-g}\left(R-{4\over3}{
(\partial\lambda)^2\over\lambda^2}+V_g(\lambda)\right)&
 \\
 S_f &= -M^3 N_f N_c \int d^5x \ V_f(\l,\t) \sqrt{-\det\left(g_{\mu\nu} + \kappa(\l)\partial_\mu \t \partial_\nu \t + w(\l) F_{\mu\nu}\right)} & 
\end{align}
are the actions for the glue and quark sectors, respectively. For the tachyon potential, we take the Ansatz $V_f(\l,\t) = V_{f0}(\l)e^{-\t^2}$. We are left with four potential functions $V_g(\l)$, $V_{f0}(\l)$, $w(\l)$, and $\kappa(\l)$ which need to be determined by qualitative and quantitative comparison to QCD data. In this article we will concentrate on chirally symmetric backgrounds describing the deconfined phase of QCD so that $\t=0$. Consequently, the results actually do not depend on $\kappa(\l)$ directly. The Ansatz for the metric can be written as
\be
 ds^2 = e^{2A(r)} \left(f(r)^{-1}dr^2 - f(r)dt^2+d\mathbf{x}^2\right) \ .
\ee

The thermodynamics in this model has been discussed at zero $\mu$ in~\cite{alte} and at nonzero $\mu$ in~\cite{altemu}, by using potentials that reproduce various features of QCD at qualitative level. It was shown that depending on the precise choice of potentials, there can be a single first order phase transition, or separate phase chiral and confinement-deconfinement phase transitions.\footnote{Notice that while ordinary QCD does not have a well defined order parameter for the deconfinement transition, there is such a parameter at large $N_c$: the pressure in the confined (deconfined) phase scales as $N_c^0$ ($N_c^2$).} In the latter case, the critical temperature of the (second order) chiral transition was found to be larger than that of the (first order) deconfinement transition in the regime of low $\mu$. In other words, an intermediate deconfined but chirally broken phase was found. At high enough $\mu$, only the deconfined chirally symmetric quark matter phase was present for all potentials. A similar phase structure has been found also in models based on a D3/D7 brane system~\cite{Evans:2010iy,Evans:2010hi}.

In the next section, we will compare the predictions of the V-QCD model to lattice data at zero chemical potential. This comparison strongly favors potentials producing phase diagrams with no intermediate (deconfined but chirally broken) phase. In the absence of the intermediate phase, the model will only have nontrivial thermodynamics in the deconfined phase.\footnote{The pressure in the confined phase is dominated by string loop contributions which have been modeled in a simple approximation in~\cite{Alho:2015zua}.} We will concentrate on analysing the results in this phase in the rest of the article.

\section{Comparison to lattice data}\label{sec:fit}

We then proceed to the determination of the potentials. The main idea is as follows. For each of the potentials, there are various qualitative constraints which restrain the dependence of the potentials on $\la$ in the UV ($\l \to 0$) and/or in the IR ($\l \to \infty$). We choose Ans\"atze which satisfy these qualitative constraints, and also include extra parameters which can then be fitted to lattice Yang-Mills and QCD data at intermediate values of the coupling. We proceed by first fitting the potential $V_g$ of the glue action to lattice data at large $N_c$. Then, keeping $V_g$ fixed, we fit the remaining potentials to QCD data at $N_c=3$ (because there is not much data available at higher $N_c$). Details will be given below and in Appendix~\ref{app:fit}. We will argue that using data at $N_c=3$ is a reasonable approach in Sec.~\ref{sec:discussion}.

In this work we will only consider lattice data for thermodynamics at $\mu=0$. Comparison to other data such as meson and glueball spectra will be considered in future work~\cite{datafit}.

Earlier related works considered fitting QCD lattice data at $\mu=0$ using a somewhat simpler holographic model based on Einstein-Maxwell-dilaton actions and studying the continuation of the equations of state to the $(\mu,T)$-plane~\cite{DeWolfe:2010he,DeWolfe:2011ts,Knaute:2017opk,Critelli:2017oub}. These articles concentrate on the physics near the critical point, whereas in the current article we will be interested in the physics at low temperatures. The main difference in the current model to the earlier work is the presence of the first order confinement-deconfinement transition even at small chemical potentials. We will discuss how the transition is treated below. 

There are also numerous other approaches which aim at modeling the thermodynamics of QCD in the deconfined phase, also taking into account constraints from lattice data. These include extensions of pQCD to relatively low temperatures and chemical potentials using quasiparticle models based on hard thermal loop computations~\cite{Rebhan:2003wn,Peshier:2005pp} and analyses based on Nambu-Jona-Lasinio model~\cite{Ratti:2005jh}. We are planning to carry out a comparison of our results for the thermodynamics to other approaches in future work.

\subsection{Glue sector}

First, the dilaton potential $V_g$ is chosen after a careful comparison to Yang-Mills data~\cite{ihqcd1,ihqcd2,ihqcd4,ihqcd5,ihqcd6} at $x_f = 0 = N_f$. We choose here a potential~\cite{Alho:2015zua} which has been shown to work well with backreacted flavors. The Ansatz reads
\be
 V_g(\lambda)=12\,\biggl[1+V_1 \l+{V_2\lambda^2
\over 1+\l/\l_0}+V_\mathrm{IR} e^{-\l_0/\l}(\l/\l_0)^{4/3}\sqrt{\log(1+\lambda/\l_0)}\biggr]\,.
\label{vg} 
\ee
This Ansatz has been chosen to reproduce qualitative features of QCD such as confinement, magnetic charge screening, and linear glueball trajectories (determining IR asymptotics) as well as asymptotic freedom (determining UV asymptotics). Here $V_1$ and $V_2$ are fixed by requiring the UV RG flow of the 't Hooft coupling to be the same as in QCD up to two-loop order. The fit parameters in the potential are then $V_\mathrm{IR}$ and $\l_0$, but also $M$ and the energy scale of the UV RG flow $\Lambda_\mathrm{UV}$ need to be determined through the fit (see Appendix~\ref{app:fit} for details).

\begin{figure}[!t]
\begin{center}
\includegraphics[width=0.8\textwidth]{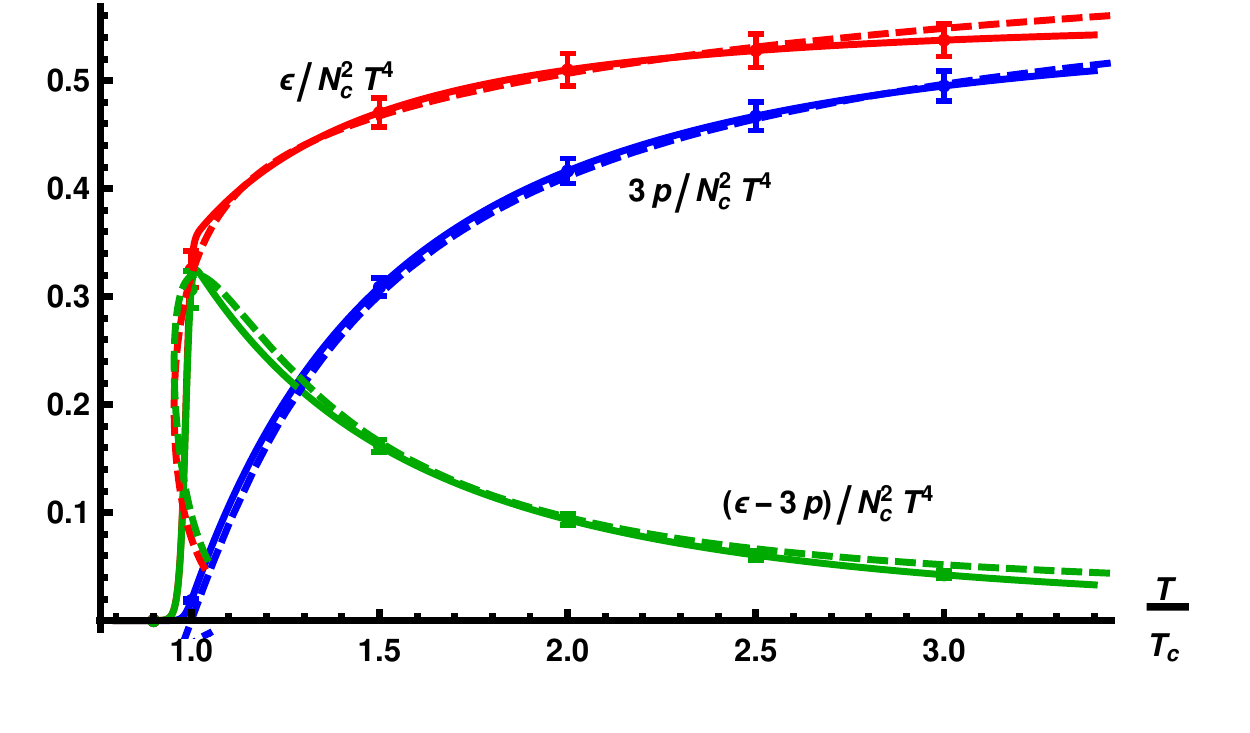}
\end{center}

\caption{Comparison of the thermodynamics of the glue sector of V-QCD to Yang-Mills lattice data~\protect\cite{Panero:2009tv}. The solid lines and error bars represent the extrapolation of lattice data to  $N_c = \infty$ and the dashed curves are the results for the holographic model. }
\label{gluefit}
\end{figure}

The result of the fitting to lattice data extrapolated to $N_c \to \infty$~\cite{Panero:2009tv} is shown in Fig.~\ref{gluefit}. Notice that the overall normalization of the pressure determines the value of $M$, the critical temperature fixes $\Lambda_\mathrm{UV}$ whereas the parameters  $V_\mathrm{IR}$ and $\l_0$ affect the shape of the curve. The fit is stiff: producing results which differ qualitatively from lattice data would require input functions which have highly nontrivial $\l$ dependence.

\subsection{Flavor sector}

The flavor sector of the model can be fitted to QCD data (now including flavors) following a rather similar strategy as for the glue sector. However since there is no lattice data available at large numbers of $N_c$ and $N_f$, we choose to compare the model to lattice data with $2+1$ flavors. 

\begin{figure}[!t]
\begin{center}
\includegraphics[width=0.49\textwidth]{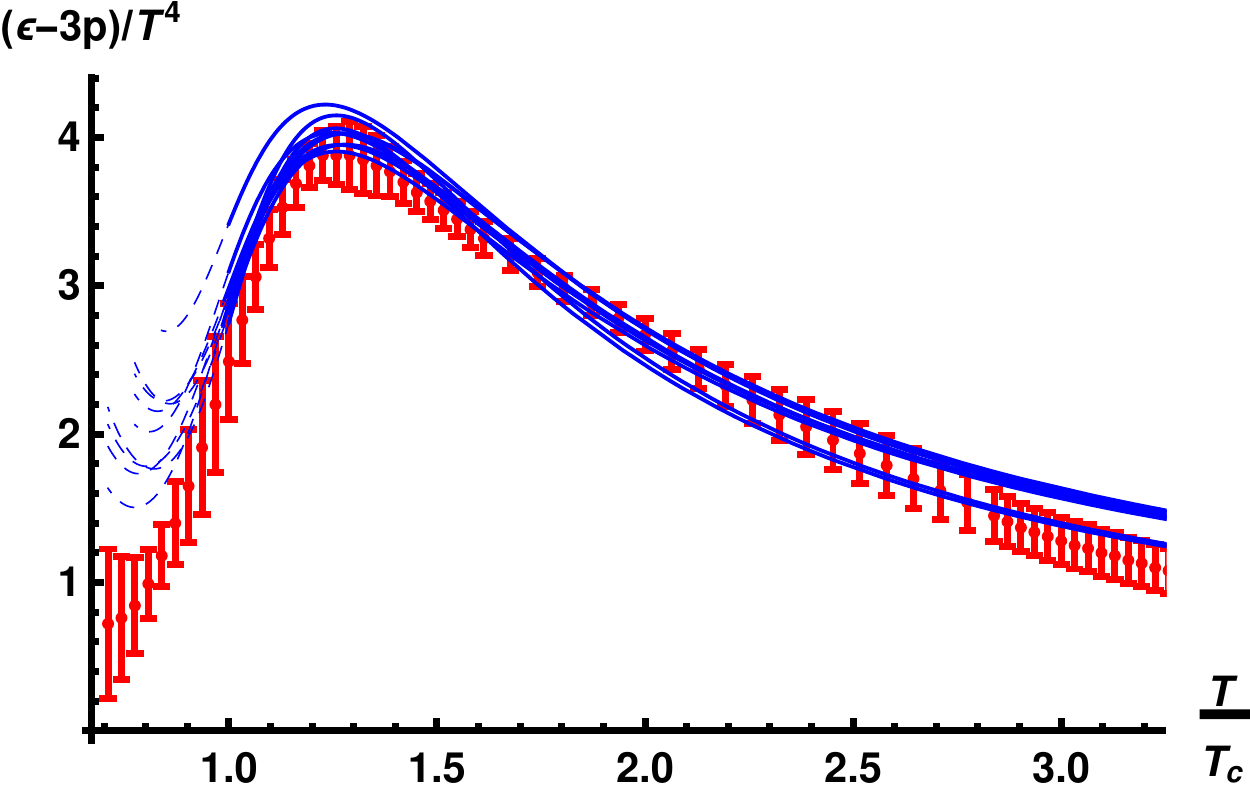}
\includegraphics[width=0.49\textwidth]{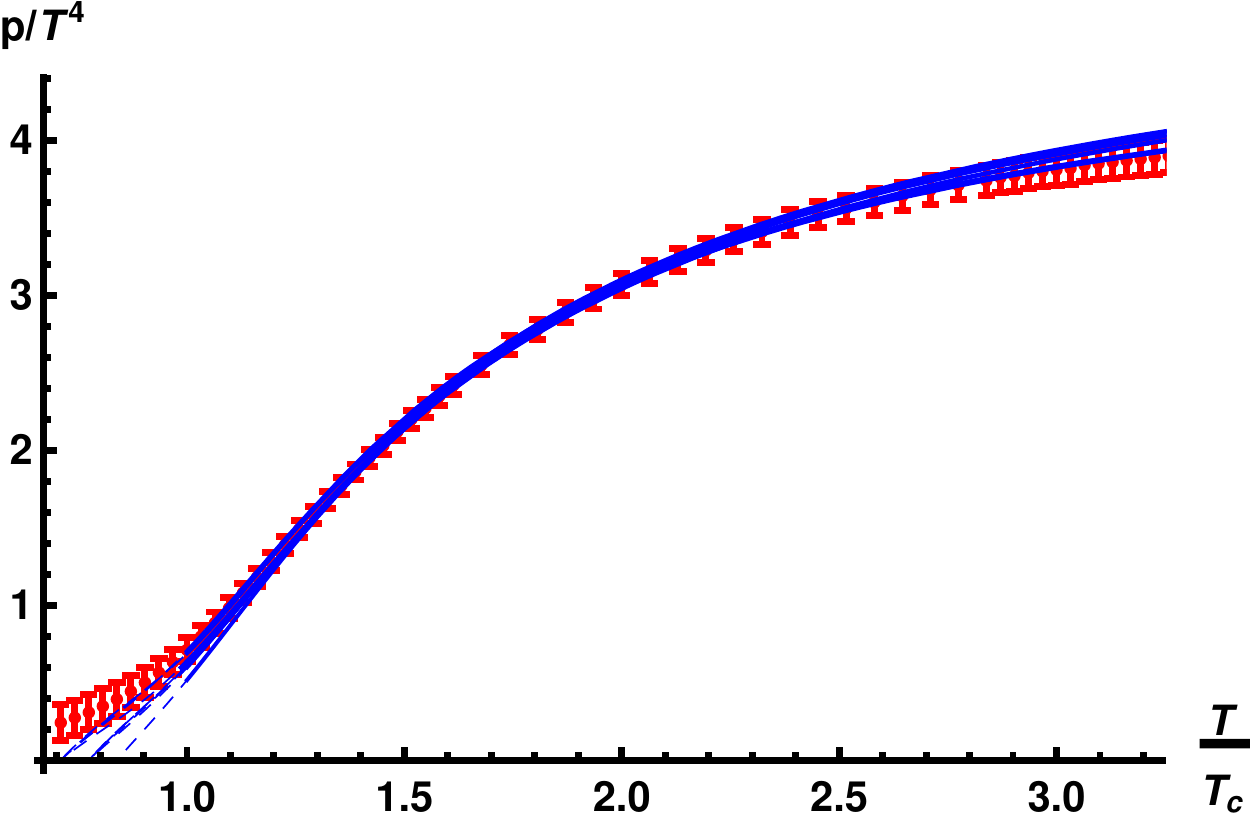}
\includegraphics[width=0.49\textwidth]{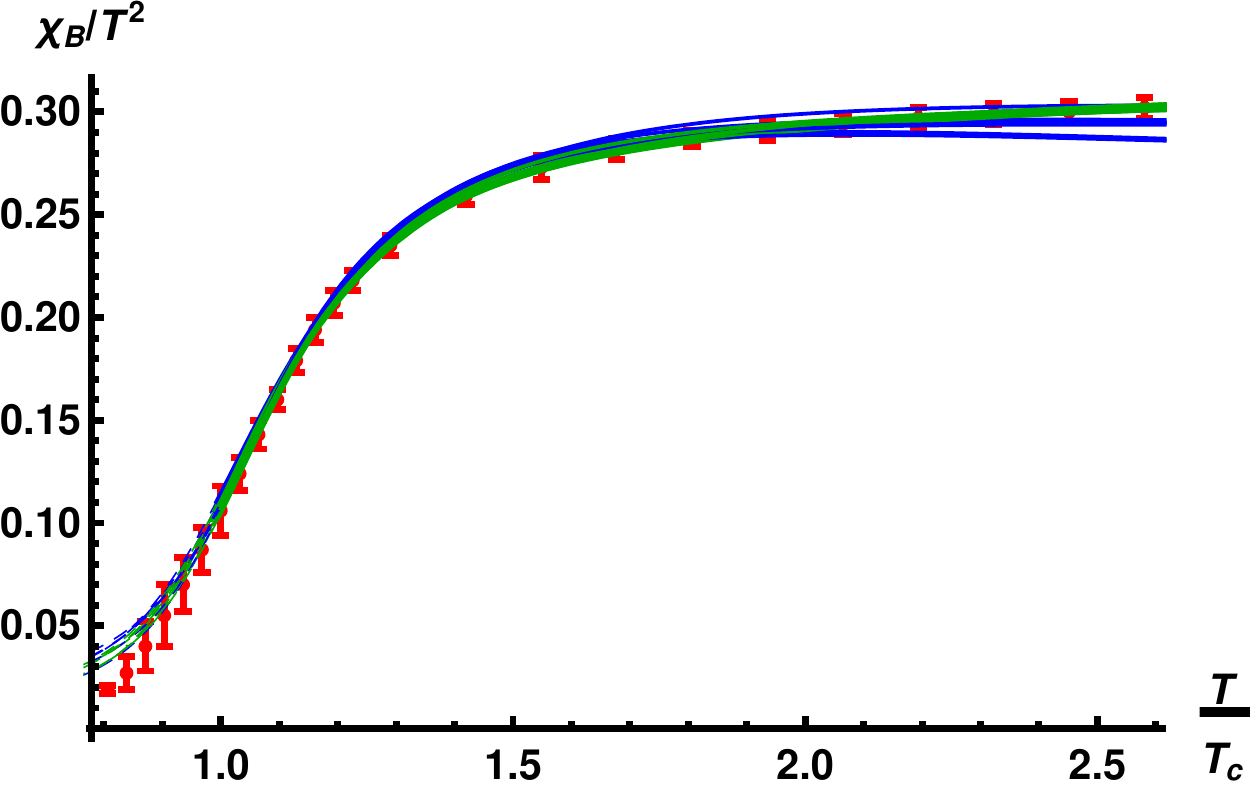}
\end{center}
\caption{Comparison of the thermodynamics of the flavor sector of V-QCD to lattice QCD data with $2+1$ flavors. The red dots and error bars are lattice data and the various curves are fit results with slightly different assumptions. }
\label{flavorfit}
\end{figure}

Our Ansatz for the flavor potentials reads
\begin{align}
V_{f0} &= W_0 + W_1 \l +\frac{W_2 \l^2}{1+\l/\l_0} + W_\mathrm{IR} e^{-\l_0/\l}(\l/\l_0)^{2} & \\
\frac{1}{w(\l)} &=  w_0\left[1 + \frac{w_1 \l/\l_0}{1+\l/\l_0} + \bar w_0 
e^{-\l_0/\l w_s}\frac{(w_s\l/\l_0)^{4/3}}{\log(1+w_s\lambda/\l_0)}\right] &\\
\frac{1}{\kappa(\l)} &= \kappa_0 \left[1+\kappa_1 \l + \bar \kappa_0 \left(1+\frac{\bar \kappa_1 \l_0}{\l} \right) e^{-\l_0/\l }\frac{(\l/\l_0)^{4/3}}{\sqrt{\log(1+\lambda/\l_0)}}\right] &
\end{align}
The asymptotics of the potentials were chosen to reproduce known features of QCD such as linear radial meson trajectories, correct behavior at large quark mass, compatibility with flavor anomaly structure, and UV properties of the flavored correlators~\cite{jk,aijk2,Jarvinen:2015ofa} (see also~\cite{Iatrakis:2014txa,Iatrakis:2015sua}). We will also consider potentials with $W_0=0$ and in this case we choose a slightly modified Ansatz for $w(\l)$, see Appendix~\ref{app:fit}.

The parameters $W_1$, $W_2$, $\kappa_0$, and $\kappa_1$ were determined by the UV dimension of the $\bar qq$ operator and the RG flow of the coupling and the quark mass in the UV. The remaining additional parameters are $W_0$, $W_\mathrm{IR}$, $w_0$, $w_1$, $\bar w_0$, $w_s$, $\bar \kappa_0$, and $\bar \kappa_1$.

These parameters are then fitted to lattice data at $\mu=0$. The fit is carried out in stages: first we consider the interaction measure and pressure of QCD, with lattice data with 2+1 dynamical flavors from~\cite{Borsanyi:2013bia}. This fit is independent of the shape of $w(\l)$, and $\kappa(\l)$ affects only the location of the phase transition of V-QCD from the confined to deconfined phase. The most relevant parameters are therefore $W_0$ and $W_\mathrm{IR}$. There is an important complication with respect to the fit to Yang-Mills data: QCD at physical quark masses has a crossover, whereas the holographic model has a first order phase transition. The analysis of~\cite{Alho:2015zua} suggests that this is due to the neglect of stringy loop contributions which are important in the confined phase. We follow the notion of the reference, and fix the phase transition temperature to around 120~MeV rather than near the peak of the susceptibilities in the lattice data around 155~MeV. The location of the transition is then expected to shift to higher temperatures and become weaker due to the aforementioned loop contributions. This strategy produces much better overall fit than fixing the transition of the holographic model to lie at, say, 155~MeV. 

We choose to work at $x_f=1$, roughly corresponding to $2+1$ flavors at $N_c=3$. We choose samples with three values for the critical temperature (110, 120, and 130~MeV) 
and four values for the parameter $W_0$ (0, 1, 2.5, and 5.886 with the last value giving the ``automatic'' normalization to Stefan-Boltzmann thermodynamics at high temperatures~\cite{alte}). The parameter $W_\mathrm{IR}$ is fitted to the functional form of the interaction measure. The parameters $\bar \kappa_0$ and $\bar \kappa_1$ are chosen to set the location of the phase transition at the desired point. Also $M$ and $\Lambda_\mathrm{UV}$ are refitted reflecting their possible dependence on $x_f$. The resulting values of the parameters can be found in Appendix~\ref{app:fit}.

We stress that the fit is again stiff and it would not be possible to find good fits to arbitrary data without introducing complicated structure in $V_{f0}$. In particular, after tuning $V_\mathrm{IR}$ such that the shape of the normalized interaction measure is roughly correct, the bump in the interaction measure is automatically reproduced by the dynamics of the model rather than put in by the fit Ansatz.

The comparison to the lattice data for basic thermodynamic functions is shown in Fig.~\ref{flavorfit} (top row). The various curves correspond to different values of the transition temperature and $W_0$ from the samples given above. The thick curves (for $T>T_c$ with $T_c =155$~MeV) represent the fit and we also show the result of the model as the dashed lines for the deconfined phase below $T_c$: as explained above, in this regime the deconfined phase should be replaced by the confined phase with added loop contributions. We plan to return to this issue in future work.

The shape of $w(\l)$ can then be fitted to the first nontrivial cumulant of the pressure at $\mu=0$, {\emph{i.e.}}, the baryon number susceptibility $\chi_B = d^2p/d\mu^2$. In our model, this can be computed at $\mu=0$ through~\cite{altemu}
\be
 \chi_B = M^3N_c^2\left[\int_0^{r_h} \frac{dr}{e^{A}\,V_f(\l,0)w(\l)^2}\right]^{-1}
\ee
where boundary was set to be at $r=0$ and $r_h$ is the location of the horizon.

For each of the $V_{f0}(\l)$ and $\kappa(\l)$ determined above we carry out two additional fits of $w(\l)$ to $\chi_B$ with lattice data from~\cite{Borsanyi:2011sw}: one rough fit with three parameters, setting $w_1=0$, and a more precise four parameter fit including also $w_1$. Notice that $w_0$ only affects the normalization of $\chi_B$. The results are shown in Fig.~\ref{flavorfit} (bottom row). The blue curves are the three parameter fits and green curves are four parameter fits. Comparison to higher order cumulants is left for future work.

\section{Holographic thermodynamics and cold QCD matter}\label{sec:thermo}

Having discussed the thermodynamics of deconfined holographic quark matter at zero chemical potential and finite temperature, we then analyse the predictions of the holographic model at zero temperature and at high baryon density. 
It is important to check that the result roughly agrees with what is already known about the QCD equation of state at zero temperature. As we will show in this section, the agreement with both nuclear matter EoS (at low densities) and pQCD (at high densities) is remarkably good, given that the only input data of the model are lattice results at $\mu =0$ and qualitative constraints.

In order to carry out the comparison, we construct a family of equations of state which interpolate between the known results from effective field theory at low densities and pQCD at high densities\cite{Kurkela:2014vha,Fraga:2015xha,Annala:2017llu}.

\subsection{Construction of the interpolated equations of state}\label{sec:polytropes}

We are interested in  quiescent neutron stars, so we focus on electrically neutral strongly interacting matter at beta equilibrium.\footnote{Notice that the quark matter phase with three massless flavors is automatically beta-equilibrated and electrically neutral in the absence of electrons.} State-of-the-art nuclear physics methods are only reliable up to around the nuclear saturation density $n_s\approx 0.16 {\mathrm{fm}}^{-3}$, whereas the densities reached at the cores of neutron stars, and where the possible quark matter phase is commonly expected to reside, are few multiples of $n_s$. This means that extrapolating even the well-established low density EoS to the regime of interest comes with huge uncertainty.

We will not choose a particular model but instead consider a whole family of EoSs, closely following \cite{Annala:2017llu} (for earlier pioneering work, see \cite{Kurkela:2014vha,Fraga:2015xha}). The key idea is that the thermodynamic functions can be arbitrary as long as the following conditions are met:
\begin{itemize}
 \item The pressure and the quark chemical potential should be continuous throughout the star and the thermodynamics has to be consistent (\emph{i.e.}, $\partial_{\mu_q} p=n_q>0,\partial^2_{\mu_q} p>0$).
 \item The speed of sound cannot exceed unity anywhere inside the star.
 \item At low density the EoS should be consistent with experimentally bound phenomenology and conform with chiral effective theory (CET). 
 \item At asymptotically high density the EoS should agree with perturbative QCD.
\end{itemize}

We first describe how to select the hadronic phase EoS. 
At low densities we use the EoS obtained from the chiral effective theory following \cite{Tews:2012fj,Hebeler:2013nza} (denoted below by HLPS), which is reliable until nuclear saturation density $n_s$. Beyond this the systematic errors will begin to grow significantly, and at $1.1n_s$ they are $24\%$. We choose this density as the starting point of the interpolation.

On the other asymptotic regime, {\emph{i.e.}}, at very high densities the EoS we employ for polytropes is the perturbative QCD one \cite{Kurkela:2009gj}. There too, there is theoretical uncertainty coming from the renormalization scale dependence. At quark chemical potential values $\mu_q\approx 0.87$GeV, the uncertainty, similarly to CET, is $24\%$. This point is then chosen as the end point of the interpolation.

The family of EoS curves that we will consider presents all the possible interpolations between the low and high density limits.
This is achieved by the following procedure. We begin by dividing the chemical potential range $[\mu_q(1.1n_s),0.87{\mathrm{GeV}}]$ into several intervals. Then on each interval we consider a polytropic EoS of the form $p=\kappa n^{\gamma}$ which are smoothly joined at the endpoints either to adjacent polytropic Ansatz or on the asymptotic regions. On each of the intervals, the parameters $\kappa$ and $\gamma$ can take arbitrary positive values as long as the requirements listed above are met. We will use  polytropic Ans\"atze with four intervals. These ``quadrutropes''
then act as our proxies for the hadronic matter EoSs and one could continue along the lines of \cite{Annala:2017llu} and implement further constraints from astrophysics to figure out what is the family of allowed EoSs 
that the dense matter could follow.

\subsection{Comparison between the holographic and interpolated equations of state}

The vantage point of holographic models is that the quark matter is strongly interacting in the cores of neutron stars, {\emph{i.e.}}, at densities which are not asymptotically high. 
Indeed, this is the regime where we believe V-QCD to be applicable. But in the regime of asymptotically high densities, the predictions of V-QCD may not be reliable. 
Therefore we do not necessarily expect the thermodynamics of V-QCD to follow the Stefan-Boltzmann (SB) behavior in the extreme densities, where the quarks are expected to become free. It has been shown, however, that for the choice of action which holographically reproduces the logarithmic UV RG flow of the coupling (and which we also use in this work) it is possible to reproduce the SB law both at infinite $T$ and infinite $\mu$~\cite{alte,altemu}. In the current article, we have chosen \emph{not} to match the normalization of the pressure to the asymptotic SB law. This was done because this matching is in slight tension with the fit to lattice data, and we have preferred the quality of the fit over the asymptotic agreement with pQCD. Consequently, the various sets of potentials constructed above will not precisely agree with pQCD at high densities. This should be kept in mind when interpreting the comparison between the interpolated and holographic EoSs which we discuss next.

\begin{figure}[!ht]
\begin{center}
\includegraphics[width=0.5\textwidth]{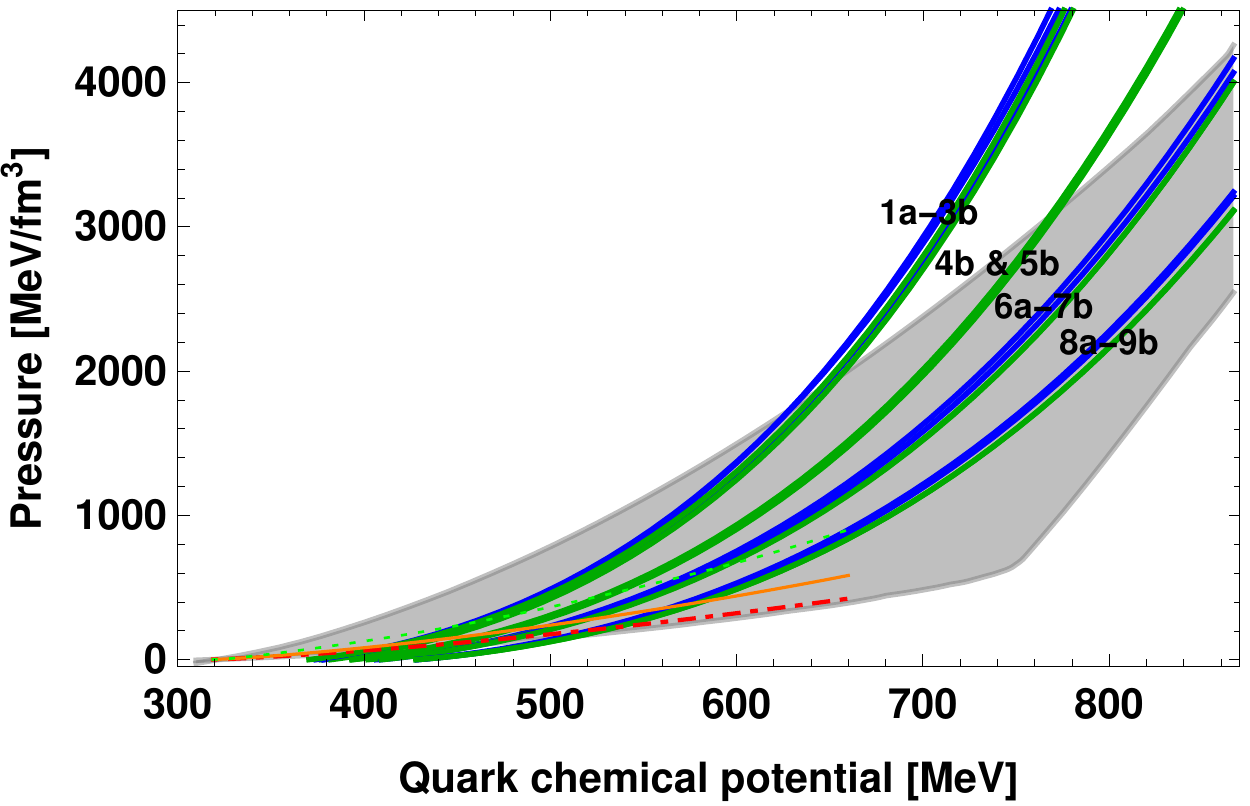}
\includegraphics[width=0.5\textwidth]{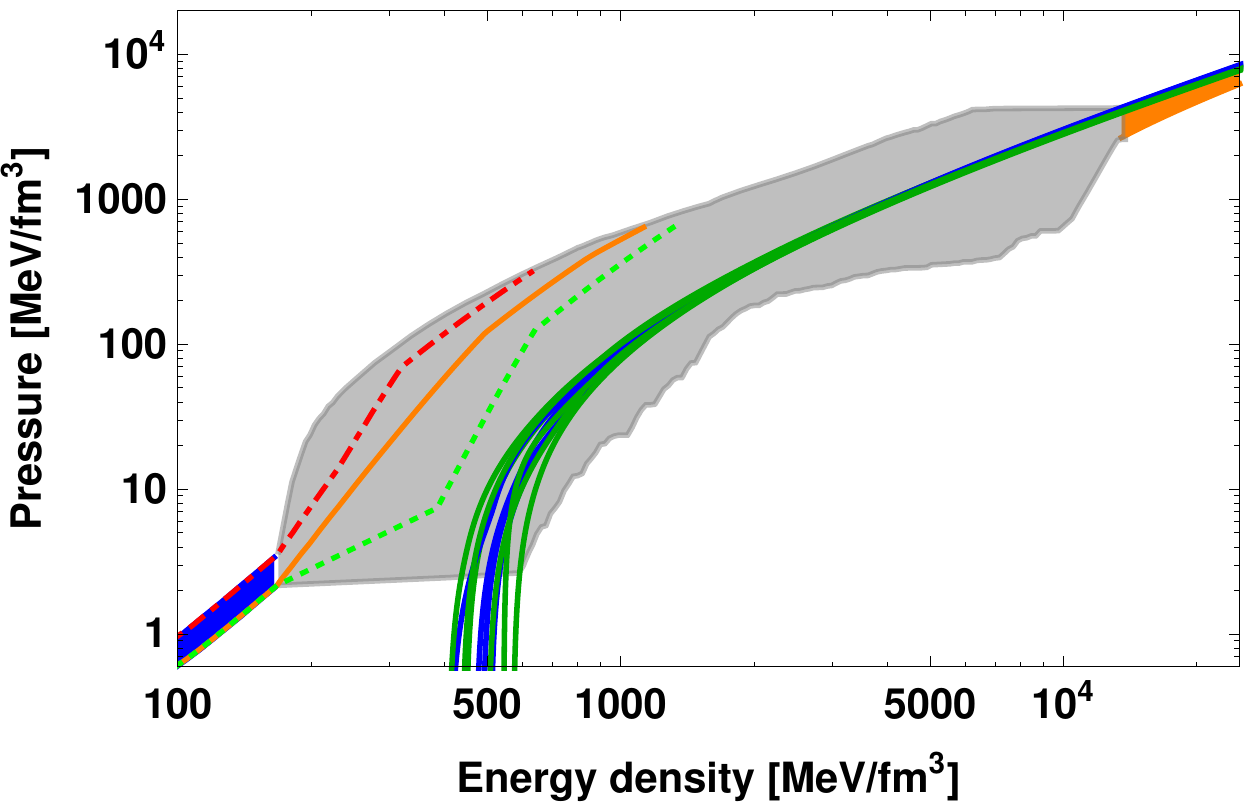}
\end{center}
\caption{Left: We plot the pressure versus the quark chemical potential for all the 16 potentials as solid green and blue curves. The labels refer to the numbering specified in Appendix~\ref{app:fit}. The three other colorful curves are determined by the HLPS equations of state \cite{Hebeler:2013nza}: the stiffest (dashed red), the softest (dashed green), and the ``typical'' one in the middle (orange). The gray band is spanned by all the interpolated EoSs described in the text.
Right: We depict the EoS in the log-log scale using the same notation as in the left hand plot. The blue and orange bands show the uncertainties of the nuclear and pQCD EoSs, respectively, in their regime of applicability.}
\label{fig:PvsCET}
\end{figure}

We present the numerical results for thermodynamics that stem from extrapolating the holographic EoS down to finite chemical potentials and to zero temperature. We depict the pressure of quark matter phase in this regime in Fig.~\ref{fig:PvsCET} (left). The different curves represent the pressures from all 16 potentials constructed in Appendix~\ref{app:fit}.  The four bunches of curves correspond to different choices for the  parameter $W_0$ (see Sec.~\ref{sec:fit} and Appendix~\ref{app:fit}). The gray band is then generated by parametrizing the low density EoS using quadrutropes as discussed in Sec.~\ref{sec:polytropes}. The parameters were chosen from a set such that we allowed a first order phase transition to occur anywhere. The bands thus represent all the physically viable EoSs. 

Remarkably, we see that the pressures from V-QCD for all choices of the potentials are in the same ballpark as extrapolated nuclear matter pressure at chemical potentials right above the vacuum-nuclear matter transition. All of the holographic curves also intersect the HLPS pressures in this regime. Moreover, the holographic curves, except for potentials {\bf{1-3}}, remain inside the gray band up to relatively high chemical potentials and pressures, where the band is already given by pQCD prediction. The agreement with pQCD is therefore good even if we did not require precise agreement between pQCD and V-QCD when choosing the potentials in Sec.~\ref{sec:fit}. Notice also that none of the V-QCD pressures are positive for $\mu_q \lesssim 370$~MeV which is well above the vacuum-nuclear matter transition. Therefore stable quark matter at low densities, and in particular quark stars, are strongly disfavored by the holographic model.

We also show the pressure as a function of energy density in Fig.~\ref{fig:PvsCET} (right) with the same notation as in the left hand plot. 
Note that in this plane the quark matter pressures do not intersect any of the HLPS curves. Therefore combining the HLPS equation of states with V-QCD necessarily leads to first order phase transitions -- we will discuss this further in the next section. All the quark matter EoSs eventually match up on the pQCD region, depicted as the orange band on the right asymptote. 

It is noteworthy that all our holographic EoSs are inside the band (apart from the regime of low pressures where they cannot represent the dominant phase as we shall demonstrate in the next section) in the pressure-energy density plane of in Fig.~\ref{fig:PvsCET} (right)  even though they deviate\footnote{The deviation for potentials {\bf{1-5}} is obvious from the plot, but actually all other potentials will also leave the pQCD band at higher values of $\mu_q$.} from the band in the left hand plot. This is possible because the mapping from one figure to another is nontrivial: in particular since $\eps = \mu_q\, dp/d\mu_q - p$, simple rescalings $\mu_q \mapsto C \mu_q$ move the curves on the left hand plot but leave the right hand plot unchanged.\footnote{Notice that the rescaling affects the functions $p(\mu_q)$ and $\eps(\mu_q)$, but leaves the EoS $\eps(p)$ unchanged.} Notice also that as we pointed out above, the parameter that affects the EoS most  is $W_0$ which creates the four families of curves in the left hand plot. At fixed $W_0$, the various choices done when fitting the lattice data are clearly a subleading effect, as can be seen from the small spread of each family of curves. The effect of varying $W_0$ is however similar to rescaling $\mu_q$: in the right hand plot most of the effect is gone and the families are no longer distinct.

Also the pressures for potentials {\bf{1a-3b}}, which have $W_0=0$, stay inside the interpolated gray band in the pressure-energy density plane. However, we 
exclude them because of their behavior on the pressure-chemical potential plane.
They fail to follow\footnote{Actually it is possible show that pressures stemming from potentials {\bf{1-3}} behave as $p\propto\mu_q ^4\log \mu_q $ at high $\mu_q$ whereas other potentials follow the SB law up to a constant.} 
the allowed band already at relative  low $\mu_q$. Therefore the existence of an equation of state which would follow the prediction from V-QCD with these potentials at intermediate $\mu_q$, \emph{i.e.}, the regime relevant for neutron stars, but still would conform to pQCD thermodynamics at extreme densities looks highly questionable.

We therefore conclude that only the EoSs for the potentials {\bf{4-9}} are thermodynamically consistent and in the correct ballpark in the sense of conforming to pQCD bound at extreme densities. However, it is immediate that some of these potentials are in tension with astrophysical constraints if a particular model from the hadronic side is chosen. This, together with finite temperature physics associated with neutron star mergers will be discussed in \cite{CheslerJokelaLoebVuorinen}. In the following we are content with discussing the EoS at zero temperature and with drawing lessons of physics interest for quiescent neutron stars.

\section{Hybrid equations of state}\label{sec:hybrids}

In this section we will discuss our results in matching the holographic equations of state to what is known at low densities or constrained from astrophysics. That is, we take the approach~\cite{Hoyos:2016zke} suggested by Fig.~\ref{fig:PvsCET} (left): we use a family of interpolated QCD EoSs to describe the low density baryonic phase, and a family of EoSs determined by V-QCD to describe the high density phase, matching the pressures in the middle where the curves intersect on the $p-\mu_q$-plane.  We aim at making generic predictions and asking if there are some universal results that can be inferred from holographic methods.

\subsection{Setup}

We compute the pressure as a function of chemical potential using our holographic framework, choosing one of the potential sets {\bf{4-9}}, and contrast it with the pressure coming from a given polytropic construction. 
Unlike in Fig.~\ref{fig:PvsCET}, we use polytropic interpolations without a first order transition, as matching between the polytrope and the holographic pressure will necessarily lead to such a transition. 
If there is a \emph{single} phase transition to the holographic quark matter phase at some quark chemical potential $\in [\mu_q(1.1n_s),0.87{\mathrm{GeV}}]$, the constructed EoS is accepted. In particular we exclude hybrid EoSs with multiple crossings between the deconfined quark matter and the hadronic matter phase. 
We then check the combinations of all holographic EoSs, using each of the potentials {\bf{4-9}}, with all the 170k interpolated polytrope EoSs, and survey the characteristic numbers for all the acceptable EoS potentials constructed this way.

To summarize, the full pressure as a function of quark chemical potential is then a piecewise hybrid curve: CET up to the point where $n_B = n_q/3$ equals $1.1n_s$, then polytropes up until a critical chemical potential $\mu_c$, where there is a first order phase transition to a holographic quark matter phase. Notice, we do not consider those polytropes that would intersect the holographic pressures at even higher chemical potentials than $0.87$GeV. We expect the pQCD approach to be valid in this regime and, moreover, at such high chemical potentials the corresponding deconfinement phase transition would occur at extremely high baryon densities ($\gtrsim 20n_s$) and would presumably leave no imprint for observers.

The last step then is to ask if such an EoS, as constructed above, will be physically plausible, given the astrophysical constraints? At current stage there are two such constraints.
The first criterion for the EoS to pass is the ability to support large enough neutron stars. This is tackled as follows. One solves the full Einstein equations for spherical configuration of a self-gravitating perfect fluid with a given energy-momentum tensor. These equations are called the Tolman-Oppenheimer-Volkov (TOV) equations and they are ordinary coupled differential equations. The solutions of these equations one obtains the mass-radius ($M-R$) relationship from where one can infer if a star of a given maximal mass is obtained (and stable). The gravitational mass of the most massive neutron star known is at least $1.97M_\odot$ \cite{Demorest:2010bx,Antoniadis:2013pzd}. Another criterion for the viability of a given EoS comes from the gravitational wave observations of merging neutron stars \cite{TheLIGOScientific:2017qsa}. An updated analysis of \cite{Abbott:2018exr} states that a neutron star of mass $1.4M_\odot$ should have a tidal deformability in the range $70 \leq \Lambda(1.4M_\odot)\leq 580$.

We therefore solve the TOV equations for the all constructed hybrid EoS as well as compute the tidal deformability for a neutron star of mass $1.4M_\odot$. If we do not find the EoS to produce a neutron star of at least of mass $1.97M_\odot$ or to satisfy the LIGO/Virgo constraint for the tidal deformability, we will discard the corresponding EoS.


\begin{figure}[!ht]
\begin{center}
\includegraphics[width=0.5\textwidth]{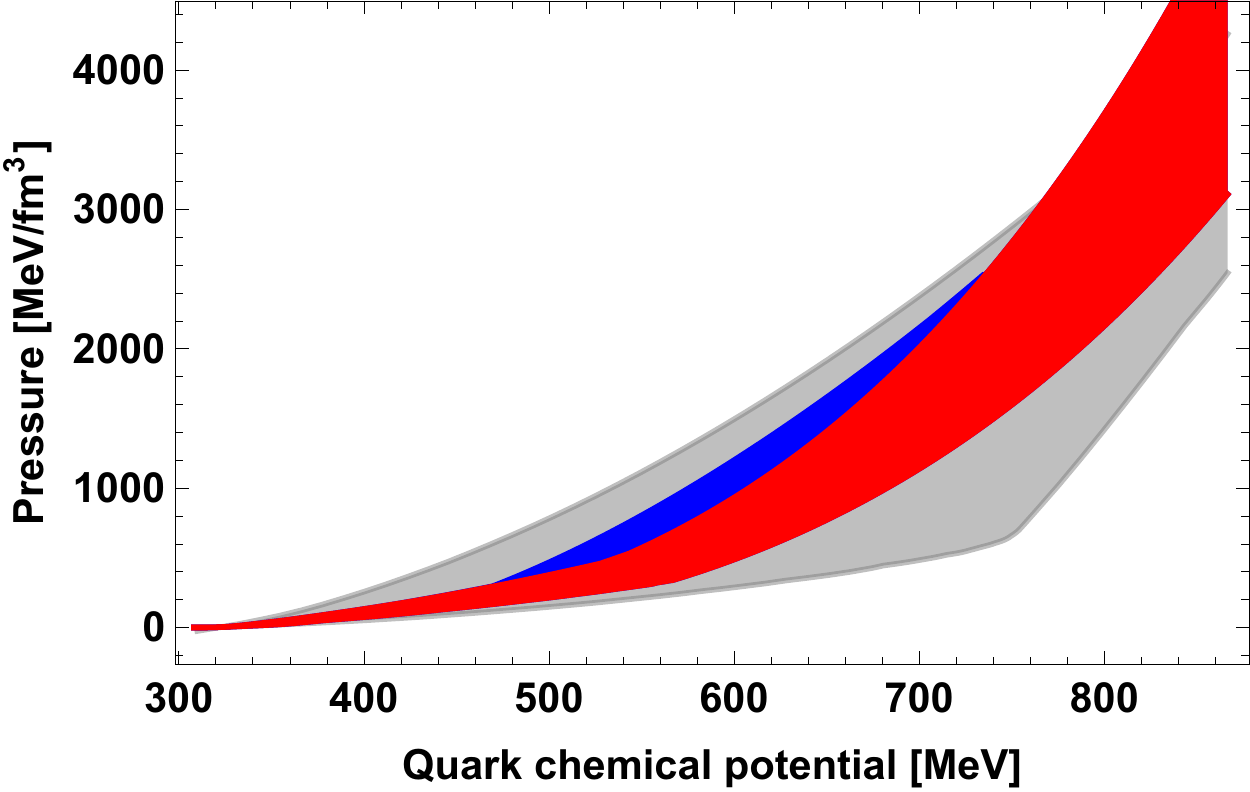}
\includegraphics[width=0.5\textwidth]{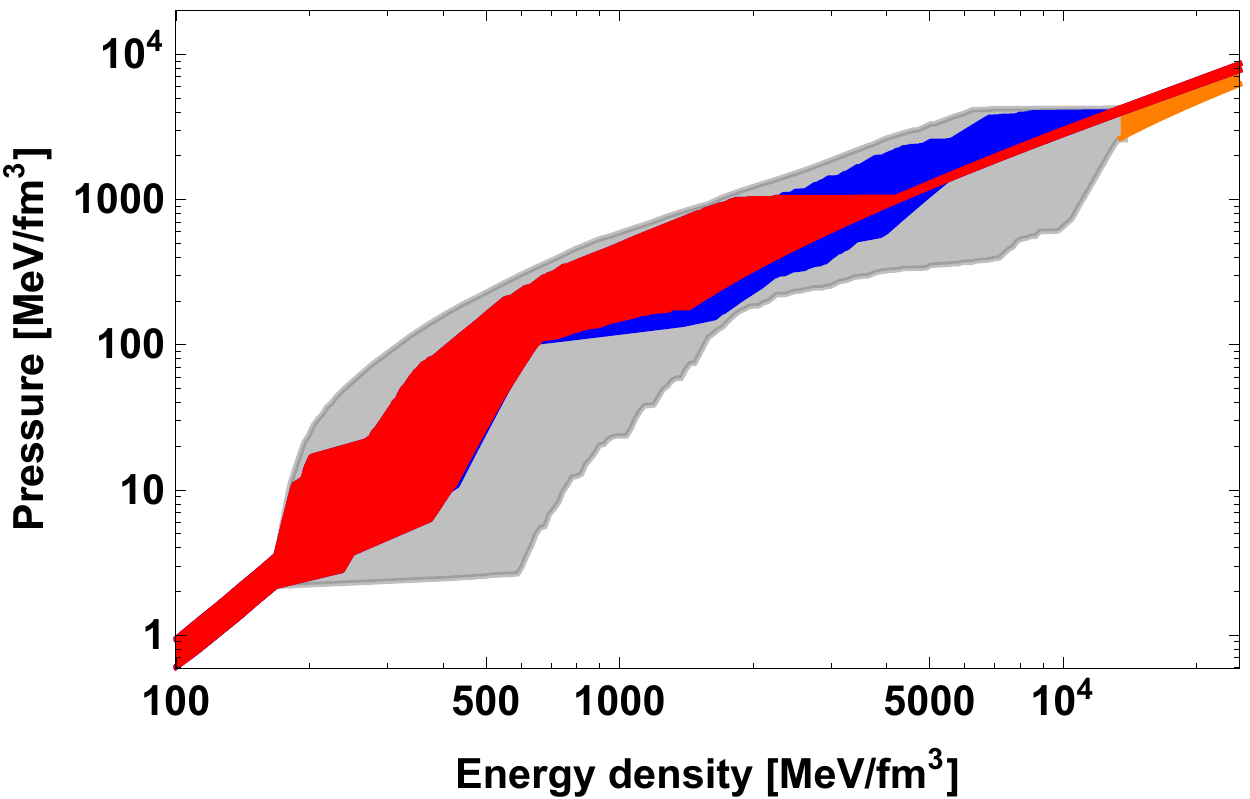}
\end{center}
\caption{The bands generated by the hybrid EoSs which survive all the criteria, including astrophysical bounds (blue), and with the additional criterion $n_B/n_s<10$ (red), compared to the band of all interpolated EoSs (gray).
Left: Bands in the $p-\mu_q$ plane. The lower edge of the blue band is not visible because it is almost identical to that of the red band. Right: Bands but in the pressure-energy density plane in the log-log scale.}
\label{fig:finalbands}
\end{figure}

\subsection{Results}

The bands spanned by the final hybrid EoSs (which also pass the astrophysical constraints) are shown in Fig.~\ref{fig:finalbands}. These bands are shown as blue and compared to the gray bands of the interpolated EoSs with a phase transition, considered in Sec.~\ref{sec:thermo}.  The holographic model restricts the band in the intermediate region between the baryonic phase and the region governed by pQCD, and in particular removes the EoSs with lowest pressures on the $p -\mu_q$ plane. On the $p-\eps$ plane, the excluded gray regions below the allowed band are due to the constraint on the maximum mass of the neutron star (low energy densities) and matching with the holographic model (high energy densities). The excluded regime above the allowed band is mostly due to the constraint on tidal deformabilities. 

It is generally believed that the baryonic phase cannot exist at densities which exceed the nuclear saturation density by more than roughly by a factor of ten as the baryons would need to be extremely tightly packed. Therefore, we have also shown the red bands where we added the additional constraint that $n_B/n_s<10$ for the hadron phase at the transition. This further reduces the bands, in particular at larger pressures, as the transitions with very high values of $\mu_q$ are excluded. Notice that the blue bands also include the red bands even in those regions of the plots where this does not seem obvious.

\begin{figure}[ht]
\begin{center}
\includegraphics[width=0.5\textwidth]{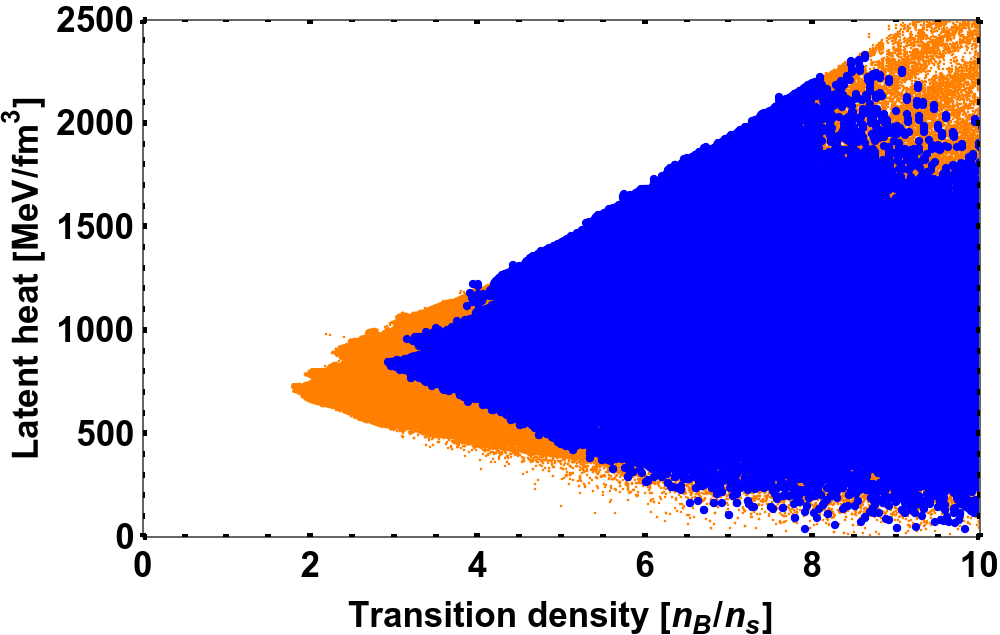}
\includegraphics[width=0.5\textwidth]{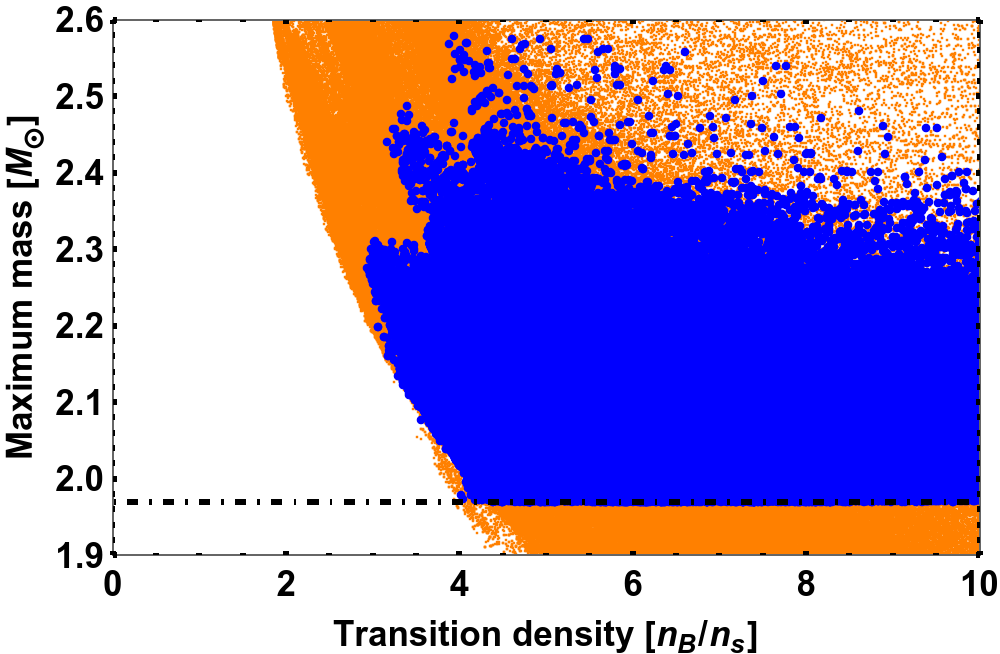}
\end{center}
\caption{Left: We plot the latent heat of the deconfinement transition versus the baryon phase transition density for all the hybrid EoSs (corresponding to potentials {\bf{4-9}} and all the quadrutropes). The blue points correspond to those EoSs that survived our criteria of sustaining massive enough neutron stars together with having tidal deformabilities in the accepted range, while the orange points fail either criterion. For $n_B/n_s \lesssim 7.5$, the latent heats are bounded both from below and above. Right: The maximum mass of the neutron star versus the transition density. The plot shows that neutron stars cannot be heavier than 2.6$M_\odot$. The horizontal dot-dashed line shows the minimal allowed value of the maximum mass.}
\label{fig:scatters}
\end{figure}

One of our most important results is shown in Fig.~\ref{fig:scatters} (left), where we surveyed all the produced EoSs and calculated the latent heats at the critical point.
Typical latent heats are sizeable, of the order 1 GeV/fm${}^3$.
The minimal transition density in the hadronic phase for the EoSs satisfying all constraints (the blue dots) is  $n_B/n_s \simeq 2.9$. 
The phase transition between the hadronic and quark matter phases at low and intermediate densities, {\emph{i.e.}}, up to $n_B/n_s \simeq 7.5$, is strongly first order. As can be verified by studying the solutions of the TOV equations for any of the EoSs in this regime, cold neutron stars cannot sustain quark matter deep in their cores due to the great energy barrier. That is, the slope of the  mass-radius curve is positive whenever the center of the star is in 
the quark matter phase. 

In the regime of high densities, \emph{i.e.}, for  $n_B/n_s \gtrsim 7.5$, we can only set a high upper limit for the latent heat. This is not surprising because as $n_B$ grows, the transition is pushed towards higher chemical potentials and the pQCD regime, where the spread of the interpolated EoSs grows and the holographic model becomes less predictive. Also weakly first order and second order transitions are possible in this region, suggesting that these EoSs can support neutron stars with stable quark matter cores. We have, however, checked the mass radius curves for selected EoSs in the range $7.5 \lesssim n_B/n_s \lesssim 10$, and quark matter cores are unstable by a clear margin for all of them.\footnote{As it turns out, a second order transition between the hadronic and quark matter phases would require a relatively sharp upwards turn in the pressure of the hadronic phase just below the phase transition. Then the neutron star actually becomes unstable already when the central pressure reaches the values near this turn, \emph{i.e.}, below the value of the hadron/quark matter phase transition.}

In Fig.~\ref{fig:scatters} (right) we have collected the data for maximum masses of the neutron stars.  We find that neutron stars cannot exceed 2.6 solar masses. There is an additional uncertainty in this plot which does not affect any of our other results: When constructing the polytropes, following~\cite{Annala:2017llu}, we chose as the  starting point either the softest or the stiffest HLPS EoS at low densities. The maximum mass of the neutron star also depends on the structure of its crust and outer core which are determined by the low density end of the EoS. Therefore adding EoSs with low density tails between the soft and stiff HLPS EoSs would cause some additional spread (amounting to less than 0.1 solar masses) of the maximum mass. Including this effect would slightly change the shape of the region with blue dots, mainly at low transition densities.

\begin{figure}[!ht]
\begin{center}
\includegraphics[width=0.5\textwidth]{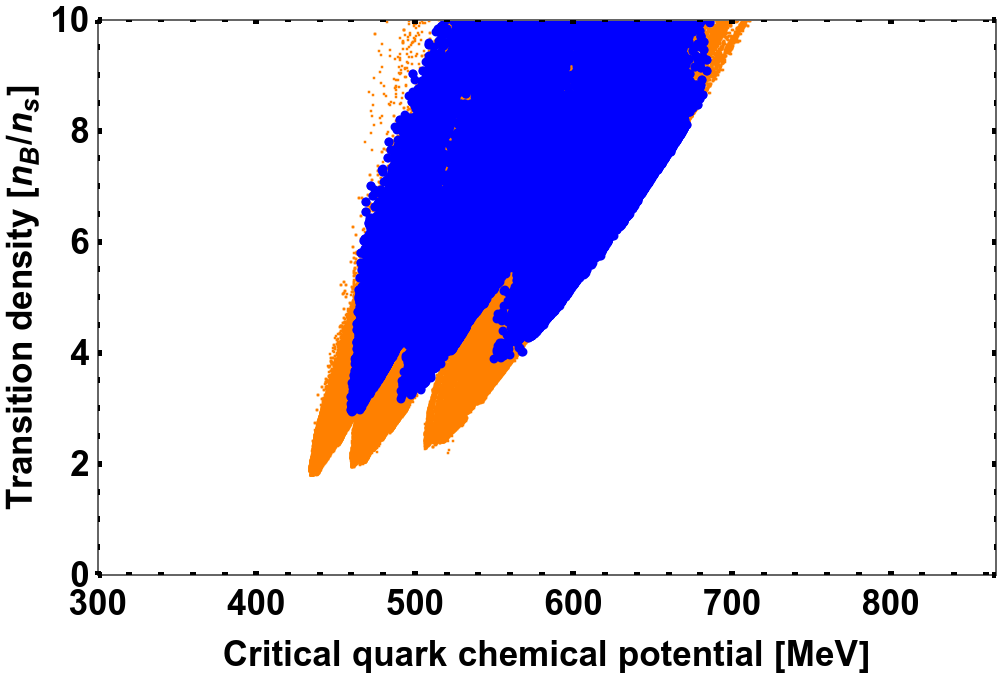}
\includegraphics[width=0.5\textwidth]{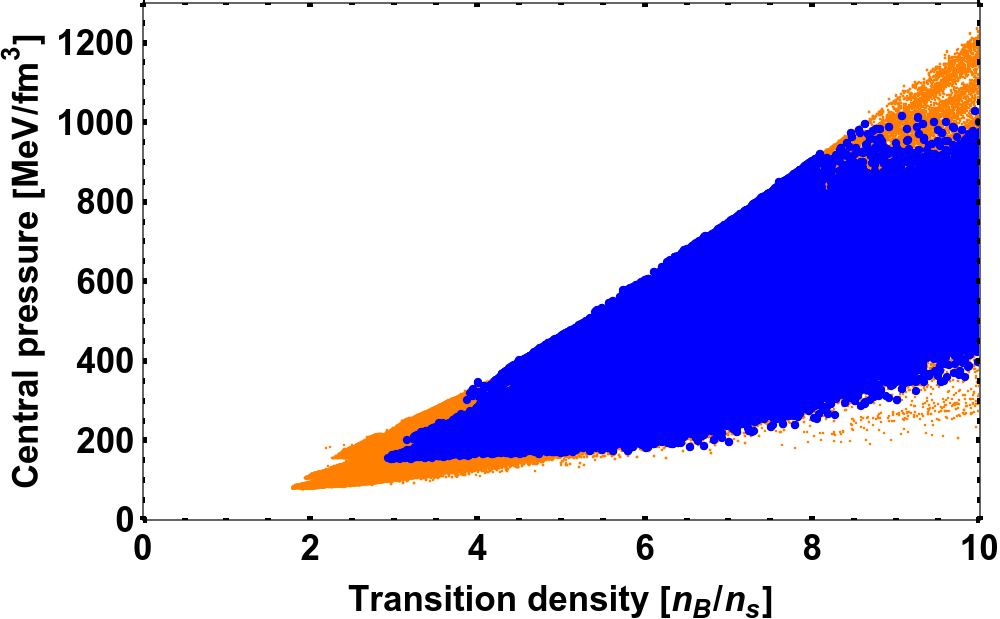}
\end{center}
\caption{Left: The transition density of the baryonic matter as a function of the critical quark chemical potential. Right: The transition pressure, or in other words the central pressure of the star as a function of baryon density.}
\label{fig:scattersnbmu}
\end{figure}

We show the transition density as a function of the critical quark chemical potential in Fig.~\ref{fig:scattersnbmu} (left). For $n_B/n_s < 10$, the critical chemical potentials are found to be between $460$~MeV and $680$~MeV. The spiky structure of the dotted regions at low densities is due to our choices of potentials in the holographic model, in particular the fact that we used potentials with three different values for the parameter $W_0$. Letting $W_0$ vary continuously would smooth out the regions and fill the spaces between the spikes. Fig.~\ref{fig:scattersnbmu} (left) shows the central pressure of the star as a function of the transition density.

\section{Discussion} \label{sec:discussion}

In this paper we took the first steps in using lattice QCD results at zero chemical potential to constrain the various scalar potentials in the V-QCD action. We demonstrated that a very good fit to the QCD pressure and its first cumulant at zero $\mu$ was possible, with errors comparable to those of the lattice data. We then projected the resulting thermodynamics in the quark matter phase to the region relevant for quiescent neutron stars, {\emph{i.e.}}, to vanishingly small temperatures and finite quark chemical potentials.

It was \emph{a priori} not at all clear how well this approach would work given that this region is well beyond the range of applicability of, \emph{e.g.}, simple minded analytic continuations of the lattice results to finite $\mu$. Despite this we found that after the fit to lattice data, the predictions of V-QCD were already tightly constrained in the region of interest. The dominant leftover freedom was seen to be the choice of the parameter $W_0$ which controls the relative normalization of the dilaton potential in the flavor term, {\emph{i.e.}}, in the tachyonic DBI action. Moreover, the resulting pressures from V-QCD for all choices of $W_0$ compared well with extrapolations of effective field theory results for baryonic matter, and as we demonstrated in Sec.~\ref{sec:hybrids}, could be used to construct ``hybrid'' equations of state which meet available constraints. This is a highly nontrivial consistency check of the model.

We then continued by analysing the results from the constructed hybrid equations of state. Once we carefully took into account the constraints coming both from the gravitational wave observations and the existence of two-solar-mass neutron stars from Shapiro delay measurements, we found that the holographic model is predictive. We found strong support to the fact that the latent heat of the transition from the hadronic to a three-flavor quark matter phase is well bounded from below at moderate baryon number densities. This then implies that the maximum mass of stable, cold,  neutron stars directly follows from the location of the deconfinement phase transition. 

It is important to bear in mind that our holographic analysis contained some uncontrolled approximations. Naturally, there is uncertainty from finite $N_c$ effects. For the glue sector, we were able to use directly lattice data extrapolated to $N_c = \infty$. For the case of full QCD, the best available data is however at $N_c=3$, and we used this data to constrain the holographic model. In the Veneziano limit, corrections at finite $N_c$ and $N_f$ are challenging to analyse on both sides of the gauge/gravity duality. To the contrary, in the case of the thermodynamics of the pure Yang Mills theory the corrections due to finite $N_c$ have been analysed on the lattice in~\cite{Panero:2009tv} and found to be small: already the results at $N_c=3$ serve as a good model for the limit $N_c \to \infty$. Therefore we expect that using $N_c=3$ lattice data for the thermodynamics of QCD in the Veneziano limit is also reasonable.

As another approximation,  we kept the quark masses of all flavors zero on the holographic side while fitting to lattice data with 2+1 quarks at their physical masses. While the perturbation due to light quark masses is expected to be insignificant compared to other uncertainties in our approach, this is not the case for the strange quark mass. Since our holographic model is effective, 
the effects due to the finite strange quark mass may however be captured by changes in the various potential functions as determined by the fit to the lattice data (and the same holds also for the effects due to finite $N_c$). While we have no way to estimate the size of the errors due to this procedure in the current setup, the fact that we obtain a reasonable model for the thermodynamics at all $\mu$ and $T$ suggests that we capture at least some of the quark mass effects without including the masses explicitly. 

The above issues motivate ways to make the holographic model increasingly realistic. Addressing the $1/N_c$ corrections would be technically demanding as it requires studying stringy corrections on the holographic side. Therefore the first step would be to add flavor dependent quark masses, and in particular to include the effect due to the strange quark mass. This would generalize the setting with  flavor independent quark masses in V-QCD which has already been studied in\cite{jk,Jarvinen:2015ofa,Arean:2016hcs}.

There are also several other possible extensions and related future projects. An ongoing work will carry out more detailed comparison of the model to QCD data~\cite{datafit}, including meson and glueball spectra, decay constants, lattice results for higher order cumulants of the pressure, etc.  It is likely that this comparison will further narrow the family of allowed holographic EoSs in the regime of low temperatures.

One should keep in mind that ``exotic'' color superconducting and color/flavor locking phase are expected for QCD at low temperatures and high chemical potentials. Such phases should also be present in the holographic model, as is suggested by the observation that the entropy does not vanish as $T \to 0$ in the relevant region~\cite{altemu}.

We have discarded some of the holographic EoSs in the current work because they did not compare well with the results from pQCD at high chemical potentials. This constraint could be made more systematic, {\emph{e.g.}}, by requiring the existence of smooth polytropic interpolation between the holographic result at intermediate $\mu$ and pQCD at high $\mu$.

Moreover, we used poorly constrained interpolated EoSs to model the baryonic phase in the current article. It may be possible to develop a reliable holographic description for this phase also. Unfortunately, holographic baryon physics is technically demanding as baryons are usually realized as soliton configurations of the gravitational dual. An ongoing work discuss baryons in V-QCD in simple approximations~\cite{Ishii:2019gta}. 
Studying magnetic effects would also be interesting, and relevant in particular to study the physics of magnetars. As a first step in this direction one could analyse the case of constant external magnetic field, which can be relatively easily added in the current holographic setup~\cite{Gursoy:2016ofp,Gursoy:2017wzz}.


\addcontentsline{toc}{section}{Acknowledgments}
\paragraph{Acknowledgments}

\noindent

We would like to thank E.~Annala, P.~Chesler, T.~Gorda, U.~G\"ursoy, T.~Ishii, E.~Kiritsis, A. Kurkela, G.~Nijs, and A.~Vuorinen for discussions. We would like to thank the authors of \cite{Annala:2017llu} for sharing with us the data that we used to generate the hadronic EoSs. This work is partially supported by the Netherlands Organisation for Scientific Research (NWO) under the VIDI grant 680-47-518 and the Delta-Institute for Theoretical Physics (D-ITP) that is funded by the Dutch Ministry of Education, Culture and Science (OCW).
N.~J. is supported in part by the Academy of Finland grant no. 1303622.

\appendix

\section{Potential Ans\"atze and fit parameters} \label{app:fit}

The basic idea is to parametrize the potentials $V_g$, $V_{f0}$, $1/w$, and $1/\kappa$ with Ans\"atze of the form
\be\label{genans}
 V(\l) = \sum_{k=0}^{n_\mathrm{UV}}V_k\ \l^k + e^{-\l_0/\l}(\l/\l_0)^\alpha (\log(1+\l/\l_0))^\beta \sum_{k=0}^{n_\mathrm{IR}}v_k\ (\l/\l_0)^{-k} 
\ee
where the first term\footnote{As we shall see below, depending on the values of $\alpha$ and $n_\mathrm{UV}$, the UV part may need to be modified slightly in order for it to be suppressed in the IR.} dominates near the UV ($\l \to 0$) and the second term which represents nonperturbative contributions, dominates in the IR ($\l \to \infty$). Potentials will tend to constants in the UV, and the powers $\alpha$ and $\beta$, which control the IR asymptotics, are determined by requiring the model to produce qualitatively different features of QCD~\cite{ihqcd2,jk,aijk2,Jarvinen:2015ofa}. Interestingly, the values of $\alpha$ found by comparing to QCD also typically match the first expectations from string theory. 

In the Ansatz~\eqref{genans}, many of the UV coefficients are determined by matching the UV dimensions of the operators and their UV RG flow with those found in perturbative QCD, and the IR coefficients remain as fit parameters which need to be compared to lattice QCD data or to experimental data for QCD. The general idea is that near the UV, at weak coupling, holographic predictions may not be reliable but we match our model there with pQCD in order to choose the best possible ``boundary conditions'' for the more interesting IR physics. In the IR, we can with the functions by comparing to some basic observables and then make predictions for more complicated observables. Relatively simple interpolations for the potentials between their expected UV and IR asymptotics produce a nice fit to the lattice QCD data. 

In addition to the expansion parameters in the potentials there are two other parameters: the 5 dimensional Planck mass $M$ and the dynamically generated energy scale $\Lambda_\mathrm{UV}$ of the background solutions. The latter may be defined in terms of the UV expansion of the dilaton:
\be \label{laUV}
 \l = -\frac{1}{b_0 \log (r\Lambda_\mathrm{UV})} -
\frac{  8 b_1 \log\left[-\log(r \Lambda_\mathrm{UV})\right]}{9 b_0^2\log(r \Lambda_\mathrm{UV})^2}+{\cal
O}\left(\frac{1}{\log(r\Lambda_\mathrm{UV})^3}\right) \ .
\ee
where $b_i$ are the coefficients of the QCD beta function in the Veneziano limit,
\be
 \beta(\l) \equiv \frac{d\l}{d\log \mu} = -b_0 \l^2 +b_1 \l^3 + \cdots
\ee 
Moreover we restrict to zero quark mass in this article, so there is no source for the tachyon field.

\subsection{Glue sector}

For the dilaton potential we choose
\be
 V_g(\lambda)=12\,\biggl[1+V_1 \l+{V_2\lambda^2
\over 1+\l/\l_0}+V_\mathrm{IR} e^{-\l_0/\l}(\l/\l_0)^{4/3}\sqrt{\log(1+\lambda/\l_0)}\biggr]\ ,
\label{vgapp} 
\ee
where requiring the holographic RG flow of~\eqref{laUV} to match with perturbative Yang-Mills theory fixes~\cite{ihqcd1}
\be
 V_1 = \frac{11}{27\pi^2} \ , \qquad V_2 = \frac{4619}{46656 \pi ^4} \ .
\ee

In order to fit the data we choose a value for $\l_0$ (which should be close to the characteristic value $4\pi^2$ in perturbation theory) and fit the remaining parameters $V_\mathrm{IR}$, $M$, and $\Lambda_\mathrm{UV}$ to lattice data. We find that the following values~\cite{Alho:2015zua} lead to a good fit, shown in Fig.~\ref{gluefit}:
\be
 \l_0 = 8\pi^2/3 \ , \qquad V_\mathrm{IR} = 2.05 \ , \qquad M^3 =  \frac{1.3}{45 \pi^2}\ , \qquad  \Lambda_\mathrm{UV} = 1.28\, T_c
\ee
where $T_c$ is the critical temperature of the lattice data. The value of $M^3$ which reproduces exactly the Stefan-Boltzmann law for the pressure at high temperatures is $1/45\pi^2$, so our fitted pressure will overshoot this limiting behavior by a factor of $1.3$.

\subsection{Flavor sector}

Let us first discuss the fitting of $V_{f0}$ and $\kappa$ to QCD lattice pressure with 2+1 dynamical flavors at zero chemical potential. As we are considering the chirally symmetric phase, the background solutions are independent of the tachyon kinetic coupling $\kappa(\l)$. Apart from the parameters of the glue sector, the entropy at $\mu=0$ therefore only depends on the function $V_{f0}(\l)$. In addition, since the tachyonic (thermal gas) vacuum is used as a reference when computing the pressure, it contains a constant which also depends on $\kappa(\l)$.

We adopt the following Ans\"atze for the functions to be fitted:
\begin{align}
 V_{f0} &= W_0 + W_1 \l +\frac{W_2 \l^2}{1+\l/\l_0} + W_\mathrm{IR} e^{-\l_0/\l}(\l/\l_0)^{2} & \\
\frac{1}{\kappa(\l)} &= \kappa_0 \left[1+ \kappa_1 \l + \bar \kappa_0 \left(1+\frac{\bar \kappa_1 \l_0}{\l} \right) e^{-\l_0/\l }\frac{(\l/\l_0)^{4/3}}{\sqrt{\log(1+\lambda/\l_0)}}\right] & \ .
\end{align}
Here requiring the UV dimension of the quark mass and the $\bar qq$ operator to match with the asymptotically free flavored theory sets
\be
 \kappa_0 = \frac{3}{2} - \frac{W_0}{8} 
\ee
for $x_f=N_f/N_c =1$.
Moreover requiring the holographic RG flow of~\eqref{laUV} to match with that of full QCD with dynamical flavors at $x_f=1$ gives
\be
 W_1 = \frac{8+3\, W_0}{9 \pi ^2} \ , \qquad W_2 = \frac{6488+999\, W_0}{15552 \pi ^4} \ ,
\ee
and requiring the RG flow of the quark mass to agree with the leading order anomalous dimension from pQCD sets
\be
 \kappa_1 = \frac{1}{3\pi^2}\ ,\quad (W_0=0) \ ; \qquad \kappa_1 = \frac{11}{24\pi^2} \ , \qquad (W_0 \ne 0) \ .
\ee

As discussed in the main text, we choose values for the parameter $W_0$ from the set $\{0,\,1,\, 2.5,\, 5.886\}$ where the last value leads to high temperature pressure at $\mu =0 $ having the correct flavor dependence (Stefan-Boltzmann law) with $x_f$-independent $M$. For the temperature of the first order phase transition in the holographic model we pick values from the set $\{110,\,120,\,130\}$~MeV, therefore fixing the physical energy scale. We keep $\l_0$ fixed to the value preferred by the Yang-Mills fit. For each pair of these reference values we then fit the parameters $W_\mathrm{IR}$, again the normalization $M$, and either $\bar \kappa_0$ or $\bar \kappa_1$ as follows. 

The decoupling of the tachyon field corresponds to the annihilation of the $D4$ and $\overline{D4}$ branes in the IR (at zero temperature) is required for correct physics such as reasonable meson spectra and correct anomaly structure~\cite{ckp}. For our choice of potentials, this sets a constraint $\bar\kappa_0 > 1.295$~\cite{jk,aijk2}. Even though no issues appear in the thermodynamic analysis of this article even if this bound is violated, we choose to satisfy it. 
Already values of $\bar\kappa_0$ very close to the lower bound lead to clear changes in the meson spectrum -- therefore we require $\bar\kappa_0>1.5$.
We fit the data taking $\bar \kappa_1=0$ if this is possible with $\bar \kappa_0>1.5$. If this is not the case, we set $\bar\kappa_0=1.5$ and fit $\bar\kappa_1$ to the data. Notice that $\kappa(\l)$ only affects the thermodynamics through the constant reference pressure of the confined phase, {\emph{i.e.}}, the thermal gas background which is independent of $T$ and $\mu$, so that fitting more than one parameter of $\kappa(\l)$ to data does not make sense.

\begin{table}
\begin{center}
\begin{tabular}{|c||c|c|c|c|c|c|c|}
\hline
& $W_0$ & $T_c$/MeV & $\Lambda_\mathrm{UV}$/MeV & $W_\mathrm{IR}$ & $\bar\kappa_0$  & $\bar\kappa_1$ & $45 \pi^2 M^3\ell^3/(1+7/4)$\\
\hline
\hline
1 & 0 & 110 & 230 & 0.8 & 1.5 & -1.09 & 1.32 \\
\hline
2 & 0 & 120 & 229 & 0.8 & 1.5 & -1.0 & 1.32 \\
\hline
3 & 0 & 130 & 230 & 0.8 & 1.5 & -0.9 & 1.34 \\
\hline
4 & 1 & 110 & 221 & 0.85 & 1.5 & -0.9 & 1.32 \\
\hline
5 & 1 & 120 & 226 & 0.85 & 1.5 & -0.8 & 1.34 \\
\hline
6 & 2.5 & 110 & 208 & 0.9 & 1.5 & -0.189 & 1.32 \\
\hline
7 & 2.5 & 120 & 211 & 0.9 & 1.5 & -0.047 & 1.32 \\
\hline
8 & 5.886 & 110 & 157 & 1.0 & 3.029 & 0 & 1.22 \\
\hline
9 & 5.886 & 120 & 158 & 1.0 & 3.35 & 0 & 1.22 \\
\hline
\end{tabular}
\end{center}
\caption{Fit to thermodynamics at $\mu=0$: values of various parameters. Here $\ell$ is the UV AdS radius and $M^3$ was normalized to the value giving the Stefan-Boltzmann law for the pressure at high temperatures. } \label{tab:thermofit}
\end{table}

The fit parameters are given in Table~\ref{tab:thermofit}. The fit was carried out by changing the values of the parameters ``by hand'' and comparing the result visually to the lattice results, see Fig.~\ref{flavorfit} (top). For the nonzero values of $W_0$ we excluded the largest reference value of $T_c$ because the quality of these fits was not high enough.   The combination $45 \pi^2 M^3\ell^3/(1+7/4)$ in the last column comes from the normalization with respect to the Stefan-Boltzmann pressure:  the listed numerical values indicate how large the pressure is at asymptotically high $T$ compared to the Stefan-Boltzmann (\emph{i.e.}, pQCD) value.

\begin{table}
\begin{center}
\begin{tabular}{|c||c|c|c|c|}
\hline
& $w_0$ & 3$w_s$ & $w_1$ & $\bar w_0$ \\
\hline
\hline
1a & 1.22 & 0.99 & 0 & 25   \\
\hline
1b&0.98& 0.88&  0.9& 37\\ 
\hline
2a&1.2& 1.1& 0& 18\\ 
\hline
2b&1.07& 1.0& 0.5& 24\\ 
\hline
3a&1.23& 1.05& 0&   19\\ 
\hline
3b&1.01& 0.91& 0.8& 33\\ 
\hline
4b&0.565& 1.01& 3.0& 55\\ 
\hline
5b&0.57& 0.94& 3.0& 65\\ 
\hline
\end{tabular}\hspace{1cm}
\begin{tabular}{|c||c|c|c|c|}
\hline
& $w_0$ & 3$w_s$ & $w_1$ & $\bar w_0$ \\
\hline
\hline
6a&1.28& 1.2& 0& 18\\ 
\hline
6b&0.875& 0.95& 1.7& 45\\ 
\hline
7a&1.28& 1.18& 0& 18\\ 
\hline
7b&0.83& 0.925& 2.0& 45\\ 
\hline
8a&1.37& 1.3& 0& 15\\ 
\hline
8b&1.09& 1.16& 1.0& 22\\
\hline
9a&1.365& 1.38& 0& 12\\ 
\hline
9b&1.05& 1.1& 1.2& 25\\
\hline
\end{tabular}
\end{center}
\caption{Fit to the baryon number susceptibility at $\mu=0$: values of the parameters in the function $w(\la)$. } \label{tab:suskisfit}
\end{table}

We then continue to fit $w(\l)$ to lattice data for the first cumulant of pressure $\chi_B = d^2p/d\mu^2$ at $\mu=0$. We use the following Ans\"atze for this function:
\begin{align}
 \frac{1}{w(\l)} &=  w_0\left[\sqrt{\l/\l_0} + \frac{w_1 (\l/\l_0)^{3/2}}{1+\l/\l_0} + \bar w_0 
e^{-\l_0/\l w_s}\frac{(w_s\l/\l_0)^{4/3}}{\log(1+w_s\lambda/\l_0)}\right] &\qquad &(W_0=0)&\\
\frac{1}{w(\l)} &=  w_0\left[1 + \frac{w_1 \l/\l_0}{1+\l/\l_0} + \bar w_0 
e^{-\l_0/\l w_s}\frac{(w_s\l/\l_0)^{4/3}}{\log(1+w_s\lambda/\l_0)}\right] &\qquad &(W_0\ne 0)\ .&
\end{align}
In the Ansatz for $W_0 \ne 0$, we could leave out the denominator factor in the term $\propto w_1$ since the IR term will be dominant as $\l \to \infty$ even without it, but we choose to keep it for the two Ans\"atze to be as similar as possible.

The fit results are given in Table~\ref{tab:suskisfit}. For each set of parameters from Table~\ref{tab:thermofit} we carried out two fits: one setting $w_1=0$ and fitting the remaining three parameters to data (rows labeled with ``{\bf{a}}'') and the other fitting all four parameters to the data (rows labeled with ``{\bf{b}}''). We excluded the fits {\bf 4a} and {\bf 5a} to the susceptibility  because their quality was significantly worse than the listed fits.  The resulting susceptibilities are shown in Fig.~\ref{flavorfit} (bottom) with blue (green) curves for three (four) parameter fits.


\begin{thebibliography}{99}

\bibitem{CasalderreySolana:2011us}
  J.~Casalderrey-Solana, H.~Liu, D.~Mateos, K.~Rajagopal and U.~A.~Wiedemann,
  ``Gauge/String Duality, Hot QCD and Heavy Ion Collisions,''
  book:Gauge/String Duality, Hot QCD and Heavy Ion Collisions. Cambridge, UK: Cambridge University Press, 2014
  [arXiv:1101.0618 [hep-th]].

\bibitem{Brambilla:2014jmp}
  N.~Brambilla {\it et al.},
  ``QCD and Strongly Coupled Gauge Theories: Challenges and Perspectives,''
  Eur.\ Phys.\ J.\ C {\bf 74} (2014) no.10,  2981
  [arXiv:1404.3723 [hep-ph]].

  
\bibitem{Annala:2017tqz}
  E.~Annala, C.~Ecker, C.~Hoyos, N.~Jokela, D.~Rodr{\'i}guez Fern{\'a}ndez and A.~Vuorinen,
  ``Holographic compact stars meet gravitational wave constraints,''
  arXiv:1711.06244 [astro-ph.HE].
  
\bibitem{Hoyos:2016zke}
  C.~Hoyos, D.~Rodr{\'i}guez Fern{\'a}ndez, N.~Jokela and A.~Vuorinen,
  ``Holographic quark matter and neutron stars,''
  Phys.\ Rev.\ Lett.\  {\bf 117} (2016) no.3,  032501
  [arXiv:1603.02943 [hep-ph]].

\bibitem{Ecker:2017fyh}
  C.~Ecker, C.~Hoyos, N.~Jokela, D.~Rodr{\'i}guez Fern{\'a}ndez and A.~Vuorinen,
  ``Stiff phases in strongly coupled gauge theories with holographic duals,''
  JHEP {\bf 1711} (2017) 031
  [arXiv:1707.00521 [hep-th]].
  
\bibitem{Hoyos:2016cob}
  C.~Hoyos, N.~Jokela, D.~Rodr{\'i}guez Fern{\'a}ndez and A.~Vuorinen,
  ``Breaking the sound barrier in AdS/CFT,''
  Phys.\ Rev.\ D {\bf 94} (2016) no.10,  106008
  [arXiv:1609.03480 [hep-th]].


\bibitem{Bedaque:2014sqa}
  P.~Bedaque and A.~W.~Steiner,
  ``Sound velocity bound and neutron stars,''
  Phys.\ Rev.\ Lett.\  {\bf 114} (2015) no.3,  031103
  [arXiv:1408.5116 [nucl-th]].

  \bibitem{jk}
M.~J{\"a}rvinen and E.~Kiritsis, {\it {Holographic Models for QCD in the
  Veneziano Limit}},  {\em JHEP} {\bf 03} (2012) 002,
  [\href{http://arxiv.org/abs/1112.1261}{{\tt arXiv:1112.1261}}].

\bibitem{ihqcd1}
U.~G{\"u}rsoy and E.~Kiritsis, {\it {Exploring improved holographic theories for
  QCD: Part I}},  {\em JHEP} {\bf 02} (2008) 032,
  [\href{http://arxiv.org/abs/0707.1324}{{\tt arXiv:0707.1324}}].

\bibitem{ihqcd2}
U.~G{\"u}rsoy, E.~Kiritsis, and F.~Nitti, {\it {Exploring improved holographic
  theories for QCD: Part II}},  {\em JHEP} {\bf 02} (2008) 019,
  [\href{http://arxiv.org/abs/0707.1349}{{\tt arXiv:0707.1349}}].

\bibitem{ihqcd3}
U.~G{\"u}rsoy, E.~Kiritsis, L.~Mazzanti, G.~Michalogiorgakis, and F.~Nitti, {\it
  {Improved Holographic QCD}},  {\em Lect. Notes Phys.} {\bf 828} (2011)
  79--146, [\href{http://arxiv.org/abs/1006.5461}{{\tt arXiv:1006.5461}}].

\bibitem{ikp1}
I.~Iatrakis, E.~Kiritsis, and A.~Paredes, {\it {An AdS/QCD model from Sen's
  tachyon action}},  {\em Phys. Rev.} {\bf D81} (2010) 115004,
  [\href{http://arxiv.org/abs/1003.2377}{{\tt arXiv:1003.2377}}].

\bibitem{ikp2}
I.~Iatrakis, E.~Kiritsis, and A.~Paredes, {\it {An AdS/QCD model from tachyon
  condensation: II}},  {\em JHEP} {\bf 11} (2010) 123,
  [\href{http://arxiv.org/abs/1010.1364}{{\tt arXiv:1010.1364}}].


\bibitem{Bigazzi:2005md}
F.~Bigazzi, R.~Casero, A.~L. Cotrone, E.~Kiritsis, and A.~Paredes, {\it
  {Non-critical holography and four-dimensional CFT's with fundamentals}},
  {\em JHEP} {\bf 10} (2005) 012,
  [\href{http://arxiv.org/abs/hep-th/0505140}{{\tt hep-th/0505140}}].

\bibitem{ckp}
R.~Casero, E.~Kiritsis, and A.~Paredes, {\it {Chiral symmetry breaking as open
  string tachyon condensation}},  {\em Nucl. Phys.} {\bf B787} (2007) 98--134,
  [\href{http://arxiv.org/abs/hep-th/0702155}{{\tt hep-th/0702155}}].

\bibitem{aijk1}
D.~Arean, I.~Iatrakis, M.~J{\"a}rvinen, and E.~Kiritsis, {\it {V-QCD: Spectra,
  the dilaton and the S-parameter}},  {\em Phys. Lett.} {\bf B720} (2013)
  219--223, [\href{http://arxiv.org/abs/1211.6125}{{\tt arXiv:1211.6125}}].

\bibitem{aijk2}
D.~Arean, I.~Iatrakis, M.~J{\"a}rvinen, and E.~Kiritsis, {\it {The
  discontinuities of conformal transitions and mass spectra of V-QCD}},  {\em
  JHEP} {\bf 11} (2013) 068, [\href{http://arxiv.org/abs/1309.2286}{{\tt
  arXiv:1309.2286}}].
  
\bibitem{ihqcd6}
U.~G{\"u}rsoy, E.~Kiritsis, L.~Mazzanti, and F.~Nitti, {\it {Improved Holographic
  Yang-Mills at Finite Temperature: Comparison with Data}},  {\em Nucl. Phys.}
  {\bf B820} (2009) 148--177, [\href{http://arxiv.org/abs/0903.2859}{{\tt
  arXiv:0903.2859}}].

\bibitem{Annala:2017llu}
  E.~Annala, T.~Gorda, A.~Kurkela and A.~Vuorinen,
  ``Gravitational-wave constraints on the neutron-star-matter Equation of State,''
  Phys.\ Rev.\ Lett.\  {\bf 120} (2018) no.17,  172703
  [arXiv:1711.02644 [astro-ph.HE]].

  \bibitem{CheslerJokelaLoebVuorinen}
 P.~M.~Chesler, N.~Jokela, A.~Loeb, and A.~Vuorinen, {\it {Quark matter generation in neutron star mergers}}, to appear.

\bibitem{alte}
T.~Alho, M.~J{\"a}rvinen, K.~Kajantie, E.~Kiritsis, and K.~Tuominen, {\it {On
  finite-temperature holographic QCD in the Veneziano limit}},  {\em JHEP} {\bf
  01} (2013) 093, [\href{http://arxiv.org/abs/1210.4516}{{\tt
  arXiv:1210.4516}}].

\bibitem{altemu}
T.~Alho, M.~J{\"a}rvinen, K.~Kajantie, E.~Kiritsis, C.~Rosen, and K.~Tuominen,
  {\it {A holographic model for QCD in the Veneziano limit at finite
  temperature and density}},  {\em JHEP} {\bf 04} (2014) 124,
  [\href{http://arxiv.org/abs/1312.5199}{{\tt arXiv:1312.5199}}]. [Erratum:
  JHEP02,033(2015)].

\bibitem{Evans:2010iy} 
  N.~Evans, A.~Gebauer, K.~Y.~Kim and M.~Magou,
  ``Holographic Description of the Phase Diagram of a Chiral Symmetry Breaking Gauge Theory,''
  JHEP {\bf 1003}, 132 (2010)
  [arXiv:1002.1885 [hep-th]].
  
\bibitem{Evans:2010hi} 
  N.~Evans, A.~Gebauer, K.~Y.~Kim and M.~Magou,
  ``Phase diagram of the D3/D5 system in a magnetic field and a BKT transition,''
  Phys.\ Lett.\ B {\bf 698}, 91 (2011)
  [arXiv:1003.2694 [hep-th]].
  
\bibitem{Alho:2015zua}
T.~Alho, M.~J{\"a}rvinen, K.~Kajantie, E.~Kiritsis, and K.~Tuominen, {\it
  {Quantum and stringy corrections to the equation of state of holographic QCD
  matter and the nature of the chiral transition}},  {\em Phys. Rev.} {\bf D91}
  (2015), no.~5 055017, [\href{http://arxiv.org/abs/1501.06379}{{\tt
  arXiv:1501.06379}}].

\bibitem{datafit}
 M.~J\"arvinen and G.~Nijs,
 work in progress.
  
\bibitem{DeWolfe:2010he} 
  O.~DeWolfe, S.~S.~Gubser and C.~Rosen,
  ``A holographic critical point,''
  Phys.\ Rev.\ D {\bf 83}, 086005 (2011)
  [arXiv:1012.1864 [hep-th]].
  
\bibitem{DeWolfe:2011ts} 
  O.~DeWolfe, S.~S.~Gubser and C.~Rosen,
  ``Dynamic critical phenomena at a holographic critical point,''
  Phys.\ Rev.\ D {\bf 84}, 126014 (2011)
  [arXiv:1108.2029 [hep-th]].

\bibitem{Knaute:2017opk} 
  J.~Knaute, R.~Yaresko and B.~K\"ampfer,
  ``Holographic QCD phase diagram with critical point from Einstein–Maxwell-dilaton dynamics,''
  Phys.\ Lett.\ B {\bf 778}, 419 (2018)
  [arXiv:1702.06731 [hep-ph]].
  
\bibitem{Critelli:2017oub} 
  R.~Critelli, J.~Noronha, J.~Noronha-Hostler, I.~Portillo, C.~Ratti and R.~Rougemont,
  ``Critical point in the phase diagram of primordial quark-gluon matter from black hole physics,''
  Phys.\ Rev.\ D {\bf 96}, no. 9, 096026 (2017)
  [arXiv:1706.00455 [nucl-th]].
  
\bibitem{Rebhan:2003wn} 
  A.~Rebhan and P.~Romatschke,
  ``HTL quasiparticle models of deconfined QCD at finite chemical potential,''
  Phys.\ Rev.\ D {\bf 68}, 025022 (2003)
  [hep-ph/0304294].
  
\bibitem{Peshier:2005pp} 
  A.~Peshier and W.~Cassing,
  ``The Hot non-perturbative gluon plasma is an almost ideal colored liquid,''
  Phys.\ Rev.\ Lett.\  {\bf 94}, 172301 (2005)
  [hep-ph/0502138].
  
\bibitem{Ratti:2005jh} 
  C.~Ratti, M.~A.~Thaler and W.~Weise,
  ``Phases of QCD: Lattice thermodynamics and a field theoretical model,''
  Phys.\ Rev.\ D {\bf 73}, 014019 (2006)
  [hep-ph/0506234].
  
\bibitem{ihqcd4}
U.~G{\"u}rsoy, E.~Kiritsis, L.~Mazzanti, and F.~Nitti, {\it {Deconfinement and
  Gluon Plasma Dynamics in Improved Holographic QCD}},  {\em Phys. Rev. Lett.}
  {\bf 101} (2008) 181601, [\href{http://arxiv.org/abs/0804.0899}{{\tt
  arXiv:0804.0899}}].

\bibitem{ihqcd5}
U.~G{\"u}rsoy, E.~Kiritsis, L.~Mazzanti, and F.~Nitti, {\it {Holography and
  Thermodynamics of 5D Dilaton-gravity}},  {\em JHEP} {\bf 05} (2009) 033,
  [\href{http://arxiv.org/abs/0812.0792}{{\tt arXiv:0812.0792}}].


  \bibitem{Panero:2009tv}
M.~Panero, {\it {Thermodynamics of the QCD plasma and the large-N limit}},
  {\em Phys. Rev. Lett.} {\bf 103} (2009) 232001,
  [\href{http://arxiv.org/abs/0907.3719}{{\tt arXiv:0907.3719}}].

\bibitem{Jarvinen:2015ofa}
M.~J{\"a}rvinen, {\it {Massive holographic QCD in the Veneziano limit}},  {\em
  JHEP} {\bf 07} (2015) 033, [\href{http://arxiv.org/abs/1501.07272}{{\tt
  arXiv:1501.07272}}].

 
\bibitem{Iatrakis:2014txa}
I.~Iatrakis and I.~Zahed, {\it {Spectral Functions in V-QCD with Matter:
  Masses, Susceptibilities, Diffusion and Conductivity}},  {\em JHEP} {\bf 04}
  (2015) 080, [\href{http://arxiv.org/abs/1410.8540}{{\tt arXiv:1410.8540}}].
  

\bibitem{Iatrakis:2015sua}
I.~Iatrakis and D.~E. Kharzeev, {\it {Holographic entropy and real-time
  dynamics of quarkonium dissociation in non-Abelian plasma}},  {\em Phys.
  Rev.} {\bf D93} (2016), no.~8 086009,
  [\href{http://arxiv.org/abs/1509.08286}{{\tt arXiv:1509.08286}}].  
    
  
\bibitem{Borsanyi:2013bia} 
  S.~Borsanyi, Z.~Fodor, C.~Hoelbling, S.~D.~Katz, S.~Krieg and K.~K.~Szabo,
  ``Full result for the QCD equation of state with 2+1 flavors,''
  Phys.\ Lett.\ B {\bf 730}, 99 (2014)
  [arXiv:1309.5258 [hep-lat]].
  
\bibitem{Borsanyi:2011sw} 
  S.~Borsanyi, Z.~Fodor, S.~D.~Katz, S.~Krieg, C.~Ratti and K.~Szabo,
  ``Fluctuations of conserved charges at finite temperature from lattice QCD,''
  JHEP {\bf 1201}, 138 (2012)
  [arXiv:1112.4416 [hep-lat]].
  
\bibitem{Kurkela:2014vha}
  A.~Kurkela, E.~S.~Fraga, J.~Schaffner-Bielich and A.~Vuorinen,
  ``Constraining neutron star matter with Quantum Chromodynamics,''
  Astrophys.\ J.\  {\bf 789} (2014) 127
  [arXiv:1402.6618 [astro-ph.HE]].
  
\bibitem{Fraga:2015xha}
  E.~S.~Fraga, A.~Kurkela and A.~Vuorinen,
  ``Neutron star structure from QCD,''
  Eur.\ Phys.\ J.\ A {\bf 52} (2016) no.3,  49
  [arXiv:1508.05019 [nucl-th]].
  
  
\bibitem{Tews:2012fj}
  I.~Tews, T.~Kr\"uger, K.~Hebeler and A.~Schwenk,
  ``Neutron matter at next-to-next-to-next-to-leading order in chiral effective field theory,''
  Phys.\ Rev.\ Lett.\  {\bf 110} (2013) no.3,  032504
  [arXiv:1206.0025 [nucl-th]].

  
  
\bibitem{Hebeler:2013nza}
  K.~Hebeler, J.~M.~Lattimer, C.~J.~Pethick and A.~Schwenk,
  ``Equation of state and neutron star properties constrained by nuclear physics and observation,''
  Astrophys.\ J.\  {\bf 773} (2013) 11
  [arXiv:1303.4662 [astro-ph.SR]].
  
\bibitem{Kurkela:2009gj}
  A.~Kurkela, P.~Romatschke and A.~Vuorinen,
  ``Cold Quark Matter,''
  Phys.\ Rev.\ D {\bf 81} (2010) 105021
  [arXiv:0912.1856 [hep-ph]].

\bibitem{Demorest:2010bx}
  P.~Demorest, T.~Pennucci, S.~Ransom, M.~Roberts and J.~Hessels,
  ``Shapiro Delay Measurement of A Two Solar Mass Neutron Star,''
  Nature {\bf 467} (2010) 1081
  [arXiv:1010.5788 [astro-ph.HE]].
  
\bibitem{Antoniadis:2013pzd}
  J.~Antoniadis {\it et al.},
  ``A Massive Pulsar in a Compact Relativistic Binary,''
  Science {\bf 340} (2013) 6131
  [arXiv:1304.6875 [astro-ph.HE]].

\bibitem{TheLIGOScientific:2017qsa}
  B.~P.~Abbott {\it et al.} [LIGO Scientific and Virgo Collaborations],
  ``GW170817: Observation of Gravitational Waves from a Binary Neutron Star Inspiral,''
  Phys.\ Rev.\ Lett.\  {\bf 119} (2017) no.16,  161101
  [arXiv:1710.05832 [gr-qc]].
  
\bibitem{Abbott:2018exr}
  B.~P.~Abbott {\it et al.} [LIGO Scientific and Virgo Collaborations],
  ``GW170817: Measurements of neutron star radii and equation of state,''
  arXiv:1805.11581 [gr-qc].

\bibitem{Arean:2016hcs}
D.~Arean, I.~Iatrakis, M.~J{\"a}rvinen, and E.~Kiritsis, {\it {The CP-odd
  sector and $\theta$ dynamics in holographic QCD}},
  \href{http://arxiv.org/abs/1609.08922}{{\tt arXiv:1609.08922}}.

\bibitem{Ishii:2019gta}
  T.~Ishii, M.~J\"arvinen and G.~Nijs,
  ``Cool baryon and quark matter in holographic QCD,''
  arXiv:1903.06169 [hep-ph].
  
    
\bibitem{Gursoy:2016ofp} 
  U.~G\"ursoy, I.~Iatrakis, M.~J\"arvinen and G.~Nijs,
  ``Inverse Magnetic Catalysis from improved Holographic QCD in the Veneziano limit,''
  JHEP {\bf 1703}, 053 (2017)
  [arXiv:1611.06339 [hep-th]].

\bibitem{Gursoy:2017wzz} 
  U.~G\"ursoy, M.~J\"arvinen and G.~Nijs,
  ``Holographic QCD in the Veneziano Limit at a Finite Magnetic Field and Chemical Potential,''
  Phys.\ Rev.\ Lett.\  {\bf 120}, no. 24, 242002 (2018)
  [arXiv:1707.00872 [hep-th]].
  
\end{thebibliography}
\end{document}